\documentclass[%
reprint,
amsmath,amssymb,
aps,
prd,
]{revtex4-2}
\usepackage{graphicx}

\usepackage{amssymb} 
\usepackage{amsmath} 

\usepackage[caption=false]{subfig}

\usepackage{xcolor}
\definecolor{lightpink}{rgb}{1,.90,.90}
\definecolor{lightred}{rgb}{1,.50,.50}
\definecolor{lightblue}{rgb}{.80,.90,1}

\newlength{\myIndent}
\usepackage[bookmarks=true,colorlinks=true,citecolor=blue,linkcolor=blue,urlcolor=magenta]{hyperref}

\usepackage{soul}
\sethlcolor{yellow}

\begin{document}

\title{A phenomenological model of magnetic flux tubes in strong fields induced by axion-origin photons}

\author{Vitaliy~D.~Rusov}
\email{Corresponding author e-mail: siiis@te.net.ua}

\author{Tatyana~N.~Zelentsova}
\affiliation{%
	Department of Theoretical and Experimental Nuclear Physics, Odessa Polytechnic National University,\\1 Shevchenko ave., Odessa 65044, Ukraine
}%

\date{\today}

\begin{abstract}

We propose a phenomenological model in which empty flux tubes in strong 
magnetic fields are associated with both the conversion of axions to photons of
axion origin to generate coronal heating, and the simultaneous preservation of 
the Parker-Biermann cooling effect by converting high-energy photons from the 
radiative zone to axions from the convective zone through the O-loop 
tachocline, is the source of the rise of the magnetic flux tube from the 
tachocline to the photosphere in the form of sunspots.

The model is characterized by two free parameters, justified by existing 
theoretical and observational constraints, the axion mass 
$m_a \sim 3.2 \cdot 10^{-2}~eV$, and asymmetric dark matter (ADM) with a 
particle mass $m_{ADM} \sim 5~GeV$ gravitationally captured by the Sun. 
In this scenario, temporal modulations of the axion density are caused by 
anticorrelated 11-year modulations of ADM gravitationally captured by the Sun.

We also obtained theoretical estimates of the velocities and times in empty 
magnetic tubes in strong fields rising from the tachocline to the surface of 
the photosphere in the form of sunspots. This ensured complete agreement with 
known experimentally measured velocities and times using local 
helioseismological time-distance inversions along the ray trajectories of 
acoustic waves. This allows us to study not only the subsurface structure and 
dynamics of active regions, but also the depths of the empty magnetic flux 
tube, extending up to 200,000~km from the tachocline to the photosphere.

The proposed model links the phenomenology of axions with the magnetic 
structures of the Sun and provides testable consequences for astrophysical and
solar observations.

\end{abstract}

\maketitle

\section{Introduction} 

It is well known that helioseismological methods are well-suited for studying
the Sun's rotation and structure from helioseismic reversals, but they have 
difficulty distinguishing the effects of very strong magnetic fields from other
structural changes, and therefore cannot definitively confirm the presence of a
magnetic field in the solar tachocline exceeding $10^5 ~G$. In other words, it 
is difficult to distinguish the effects of very strong magnetic fields in the 
tachocline, but this does not mean they should be discarded due to their 
virtually undetectable nature using direct seismic methods.

This raises the question of why the dynamo action is only associated with a 
magnetic field of about $10^5 ~G$, but is definitely not associated with a 
stronger magnetic field of about $10^7 ~G$? It is known that ``...even if 
complete knowledge of the solar dynamo existed -- and it does not...'', as has
been pointed out throughout the review~\cite{Charbonneau2023} -- a number of 
crucial questions require answers, including, at least: How do sunspots with 
structure, rotation velocity, and 11-year luminosity form and disappear? 
Why has spot formation not been captured even in local models, let alone global
models? How do turbulence coefficients -- turbulent pumping, turbulent 
diffusion -- change with rotation velocity, luminosity, and internal structure?
Why the inability to model accurately enough the surface layers where the 
density drops vertiginously is, according to~\cite{Kapyla2023}, the common 
characteristic of all existing dynamo models? At that, using the so-called new
era of 3D modeling and observations~\cite{Charbonneau2023}, they (solar 
physicists), surprisingly, believe that the study of the solar dynamo will be 
definitely solved within the next 10-20 years.

But the most important thing is that any dynamo action that involves, for 
example, the 11-year oscillations of sunspots or a sunspot butterfly diagram,
must contain a small number of ``free parameters'' -- usually three or less 
(see Dahlquist's key theorem~\cite{Dahlquist1963}) -- because of the 
mathematical structure of the problem, where the independence of the equations,
the stability of the system, and the dimensionality restrictions reduce the 
number of ``degrees of freedom'' needed to determine the solution.

We could raise a few more questions, but it is enough for us that ``...not 
enough for a scientific understanding of the solar dynamo. For there is an 
unfortunate revelation: with four free parameters you can provide a 
satisfactory fit to the horizon of New York or Beijing'' -- as was said by the
great Eugene~N.~Parker, who in 2009 in his last article~\cite{Parker2009} first
began to strongly doubt the existence of the solar dynamo.

Now, in contrast to the no more than of three ``free'' parameters of the solar
dynamo, consider our next two ``free'' parameters: the dark matter axions in 
the Sun's core and the dark matter asymmetry (ADM) gravitationally captured by
the Sun. These variables in the corresponding models cannot be predicted by the
theory itself and must be determined from experimental data, as they are 
related to fundamental physical constants.

Let us first consider the first ``free parameter'' -- the dark matter axions in
the Sun's core, which are necessary to solve the well-known problem of energy 
transfer by magnetic flux tubes from the tachocline to the Sun's photosphere.

It is known that the unsolved problem of energy transport by magnetic flux
tubes at the same time represents another unsolved problem related to the
sunspot darkness (see 2.2 in~\cite{Rempel2011}). Of all the known concepts 
playing a noticeable role in understanding the connection between the energy
transport and sunspot darkness, let us consider the most significant theory,
in our view. It is based on the Parker-Biermann cooling
effect~\citep{Parker1955a,Biermann1941,Parker1979b} and originates from the
early works of Biermann~\cite{Biermann1941} and Alfv\'{e}n~\cite{Alfven1942}.

As you know, the Parker-Biermann cooling
effect~\citep{Parker1955a,Biermann1941,Parker1979b}, which plays a role in our
current understanding, originates from Biermann~\cite{Biermann1941} and 
Alfv\'{e}n~\cite{Alfven1942}: in a highly ionized plasma, the electrical
conductivity can be so large that the magnetic fields are frozen into the
plasma. Biermann realized that the magnetic field in the spots themselves can
be the cause of their coolness -- it is colder because the magnetic field
suppresses the convective heat transport. Hence, the darkness of the spot is
due to a decrease in surface brightness.

Parker \citep{Parker1955a,Parker1974a,Parker1974b,Parker1974c,Parker1979b} 
pointed out that the magnetic field can be compressed to the enormous
intensity only by reducing the gas pressure within the flux tube relative to
the pressure outside, so that the external pressure compresses the field.
The only known mechanism for reducing the internal pressure sufficiently is a
reduction of the internal temperature over several scale heights so that the
gravitational field of the Sun pulls the gas down out of the tube (as described
by the known barometric law $dp / dz = - \rho g$). Hence it appears that the
intense magnetic field of the sunspot is a direct consequence of the observed
reduced temperature~\citep{Parker1955a}.

On the other hand, Parker \citep{Parker1974c,Parker1977} has also pointed out
that the magnetic inhibition of convective heat transport beneath the sunspot,
with the associated heat accumulation below, raises the temperature in the
lower part of the field. The barometric equilibrium leads to enhanced gas
pressure upward along the magnetic field, causing the field to disperse rather
than intensify. Consequently, \cite{Parker1974c} argued that the
temperature of the gas must be influenced by something more than the inhibition
of heat transport!

Our idea is that the explanation of sunspots is based not only on the 
suppression of convective heat transfer by a strong magnetic field (of the 
order of $\sim 10^7 ~G$~\citep{RusovDarkUniverse2021}) through the enhanced 
cooling of the Parker-Biermann effect~\citep{Parker1974a}, but also on the 
appearance of the axions of photonic origin
(Fig~\mbox{\ref{fig-lampochka}}, Fig.~B.1 in \cite{RusovDarkUniverse2021}) 
from the tachocline to the photosphere, which is confirmed by the 
``disappearance'' of the heat, and consequently, the temperature in the lower
part of the magnetic tube~\citep{Parker1974c,Parker1977} due to the axions of
photonic origin from the photon-axion oscillations in the O-loop near the
tachocline (see Fig.~\mbox{\ref{fig-lampochka}}).

This means that the appearance of axions of photonic origin, with which the 
problem of temperature increase in the lower part of a strong magnetic tube
disappears, and the photons of axionic origin, which have the free path
(Rosseland length; see Fig.~B.3 in~\citep{RusovDarkUniverse2021}) from the
tachocline to the photosphere, are the explanation of sunspots based not only
on the suppression of convective heat transport by a strong magnetic
field~\citep{RusovDarkUniverse2021}, but also on the indispensable existence of
the Parker-Biermann cooling effect. At the same time, we clearly understand 
that stronger fields (e.g. of the order of $10^7 ~G$) would seriously suppress
the action of the dynamo (see~\citep{Deluca1986}).

On the other hand, we understand that the existence of the Parker-Biermann
cooling effect is associated with the so-called thermomagnetic
Ettingshausen-Nernst  effect (see Apendix~A in \cite{RusovDarkUniverse2021}).
Due to the large temperature gradient in the tachocline, the thermomagnetic EN
effect~\cite{Ettingshausen1886,Sondheimer1948,Spitzer2006,Kim1969} creates
electric currents that are inversely proportional to the strong magnetic field
of the tachocline.
It means that, using the thermomagnetic EN eﬀect, a simple estimate of the 
magnetic pressure of an ideal gas in the tachocline of e.g. the Sun,

\begin{equation}
\frac{B_{tacho}^2}{8 \pi} = p_{ext} \approx 6.5 \cdot 10^{13} \frac{erg}{cm^3} ~~
at ~~ 0.7 R_{Sun}, 
\label{eq05-001}
\end{equation}

\noindent
can indirectly prove that toroidal magnetic field of the tachocline is exactly 
equal to

\begin{equation}
B_{tacho}^{Sun} = 4.1 \cdot 10^7 ~G ,
\label{eq05-002}
\end{equation}

Using the strong MFT field, we are interested in the existence of dark matter 
axions identical to solar axions, which strongly affects the conversion of 
axion origin photons in magnetic O-loops near the tachocline, and are thus 
related to the so-called thermomagnetic Ettingshausen-Nernst effect through the
Parker-Biermann cooling effect (see Apendix~A in~\cite{RusovDarkUniverse2021}).
In order to answer this question, let us first consider all the unexpected and
intriguing implications of the 11-year modulations of the ADM density in the
solar interior and around the BH (see Sect.~3 in~\cite{RusovDarkUniverse2021}).

It is well known that virtually all physicists, and astrophysicists in 
particular, know that asymmetric dark matter (ADM) exists in the Universe, 
and of course, in our galaxy, with the highest density of ADM located precisely
at the galactic center, i.e. near the black hole. At the same time, we know 
that the periods, velocities, and modulations of the S-stars are the essential
indicator of the modulation of the ADM halo density in the fundamental plane of
the Galaxy's center. Surprisingly, between the black hole and the disk, there 
are also so-called amazing S-stars~\footnote{Let us remind from where the fairly
young S-stars (Fig.~8a
in~\cite{RusovDarkUniverse2021}) appear in our galaxy between the black hole
and the disk. Since we know that in our galaxy, in addition to the supermassive
black hole (SMBH), there is the so-called intermediate black hole (IMBH; see 
evidence~\cite{Takekawa2019}). If we take into account
the evolution of binary SMBH-IMBH, then when one body (SMBH) exchanges partners
of binary stars orbiting the IMBH (Fig.~8b in~\cite{RusovDarkUniverse2021}), 
and through the extreme gravitational-tidal field, one star is captured (SMBH) 
and loses energy, while the other escapes, receives all this energy and, due to 
the hypervelocity of B-type stars -- the IMBH sling (Fig.~8b 
in~\cite{RusovDarkUniverse2021}), simply flies out of the galaxy. From here it 
becomes clear how in other galaxies a young S-star can end up between a black 
hole and a disk.}, 
of which there are just
over 30. Most intriguingly, the closest S-star to the black hole is called 
S-102, whose orbital period around the black hole is 11 years (see Fig.~6 and 
Fig.~8 in~\cite{RusovDarkUniverse2021}). From this we understand that when the
S-102 passes very close to the black hole, the high speed of this star is such
that dark matter is practically not captured by this star. But when the star 
S-102 orbits farther from the black hole, the star gravitationally captures 
these ADMs. If the modulations of the ADM halo at the GC lead to modulations of
the ADM density on the surface of the Sun (through vertical density waves from
the disk to the solar neighborhood), then there is an "experimental"
anticorrelation identity between such indicators as the ADM density modulation
in the solar interior and the sunspot cycles. Or equivalently, between the
modulation of solar axions (or photons of axion origin) and the sunspot
cycles! 

A hypothetical pseudoscalar particle called axion is predicted by the theory
related to solving the CP-invariance violation problem in QCD. The most
important parameter determining the axion properties is the energy scale $f_a$
of the so-called U(1) Peccei-Quinn symmetry violation. It determines both the
axion mass and the strength of its coupling to fermions and gauge bosons
including photons. However, in spite of the numerous direct experiments, axions
have not been discovered so far. Meanwhile, these experiments together with the
astrophysical and cosmological limitations leave a rather narrow band for the
permissible parameters of invisible axion (e.g.
$10^{-6} eV \leqslant m_a \leqslant 10^{-2} eV$ (see~\citep{ref01,ref02}; 
Fig.~13 in~\cite{CaputoRaffelt2024}, Fig.~24 in~\cite{OHare2024})).
The PQ mechanism, solving the strong CP problem in a very elegant 
way~\citep{PecceiQuinn1977,PecceiQuinn1977PRD,Wilczek1978,Weinberg1978}, is
especially attractive here, since the axion is also a candidate for
dark matter~\citep{Preskill1983,Abbott1983,Dine1983}.

\begin{figure*}
\noindent
  \begin{center}
    \includegraphics[width=16cm]{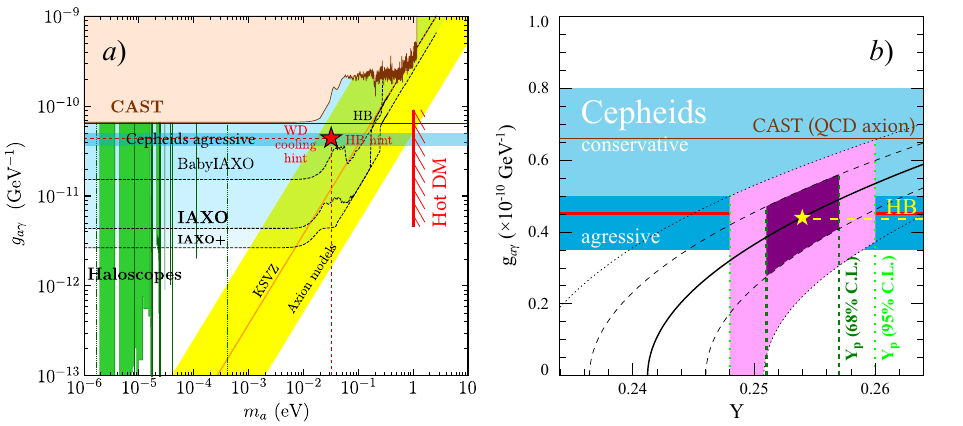}
  \end{center}
\caption{(a) Summary of astrophysical, cosmological and laboratory constraints
on axions. Comprehensive axion parameter space, highlighting two main front
lines of direct detection experiments: helioscopes (CAST~\cite{CAST2017}) and
haloscopes (ADMX~\cite{Asztalos2010}, and RBF~\cite{Wuensch1989} and microwave resonators~\cite{Brubaker2017,Zhong2018,Braine2020,Lee2020,Backes2021}).
The astrophysical bounds from horizontal branch and massive stars are labeled
``HB''~\cite{Raffelt2008} and ``Cepheids''~\cite{Carosi2013}, respectively, and
there are also astrophysical hints (WD cooling hints, and HB hint). 
The QCD motivated models (KSVZ~\cite{Kim1979,Shifman1980}) for axions lay in
the yellow diagonal band. A plot of $g_{a\gamma}$ versus $m_a$ with the most
stringent results (solid lines) and sensitivity perspectives (dashed lines) of
observations and experiments directly comparable to the different phases of
IAXO are shown, BabyIAXO, IAXO, and an upgraded version of IAXO, IAXO+. 
The yellow band denotes the region of the parameter space favoured by QCD axion
models. The red star marks the values of the axion mass 
$m_a \sim 3.2 \cdot 10^{-2}~eV$ and the axion-photon coupling constant 
$g_{a\gamma} \sim 4.4\cdot 10^{-11}~GeV^{-1}$, which
were first obtained experimentally (see~\cite{RusovDarkUniverse2021}).
(b) $R$ parameter constraints, which compares the numbers of stars in the the
horizontal branch (HB) and in the upper portion of the red giant branch (RGB),
to helium mass fraction $Y$ and axion coupling $g_{a\gamma}$ (adopted 
from~\cite{Ayala2014}). The resulting bound on the axion 
($g_{10} = g_{a\gamma} / (10^{-10}~GeV^{-1})$ is somewhere between rather 
conservative $0.5 < g_{10} < 0.8$ and most aggressive 
$0.35 < g_{10} < 0.5$~\cite{Friedland2013}. 
The red line marks the value of the axion–photon coupling constant 
$g_{a\gamma} \sim 4.5\cdot 10^{-11}~GeV^{-1}$ adopted from~Eq.~(2) 
in~\cite{Ayala2014}. The blue shaded area represents the bounds from Cepheids
observation. The yellow star corresponds to $Y=0.254$ and the bounds from HB
lifetime (yellow dashed line).}
\label{fig-axion-constraints}
\end{figure*}

At present, we have shown that axions born in the core of the Sun, and photons
of axion origin (in the million-degree solar corona)
yield a tight constraint on the axion-photon coupling constant 
$g_{a\gamma} \sim 4.4\cdot 10^{-11}~GeV^{-1}$ and mass 
$m_a \sim 3.2 \cdot 10^{-2}~eV$ (see Fig.~\ref{fig-axion-constraints}a; also 
Fig.1 in~\cite{RusovDarkUniverse2021}). Most importantly, hadronic axions 
produced in the solar core are controlled by anticorrelated 11-year variations
in the density of asymmetric dark matter (ADM) gravitationally trapped in the 
solar interior (via vertical density waves from the black hole disk to the 
solar neighborhood), with the hard part of the solar photon spectrum being of 
axion origin, which is defined as the product of the axions fraction in the 
solar core and the fraction of the sunspot area. The latter is the result of 
solar luminosity variations (see Eqs.~(27)-(28) 
in~\cite{RusovDarkUniverse2021}). This means that solar luminosity variations 
and other solar cycles are the result of 11-year variations in ADM density, 
which are anticorrelated with the density of hadronic axions in the solar core.

Interestingly, in contrast to our Fig.~\ref{fig-axion-constraints}a, in 
Fig.~\ref{fig-axion-constraints}b, we can see a well-known strong constraint on
the axion-photon coupling $g_{a\gamma} \sim 4.4\cdot 10^{-11}~GeV^{-1}$, but in
a wide mass range (see Fig.~\ref{fig-axion-constraints}b and Eq.~(2) 
in~\citep{Ayala2014}) from an analysis of a sample of 39 Galactic Globular 
Clusters. However, in our case (Fig.~\ref{fig-axion-constraints}a) the wide 
range of masses~\citep{Ayala2014} is, oddly enough, replaced by a single mass
$m_a \sim 3.2 \cdot 10^{-2}~eV$, which fits not only the axions of galactic 
globular clusters~\citep{Ayala2014}, but also all axion stars in our galaxy, 
including neutron stars~\citep{Buschmann2021}, 
supernovae~\citep{Carenza2019,Carenza2021}, and the 
Sun~\citep{RusovDarkUniverse2021}.

Taking into account the above, we structure our paper as follows. 
In Section~\ref{sec-empty-tubes}, we discuss the physics of nearly empty flux 
tubes and the connection with dark matter axions out of the solar core. 
In Section~\ref{sec-heating}, we present convective heating and magnetic 
buoyancy of flux tubes by means of axion origin photons. 
In Section~\ref{sec-reconnection}, we discuss the physics of primary and 
secondary reconnection of flux tubes in the lower layers and the features of 
the Joy's law tilt angle. 
In Section~\ref{sec-double-maxima}, we discuss the physics of the observed 
double maxima of sunspot cycles, where the first maximum does not depend on 
the tilt angle of the Joy's law, while the second one definitely depends on the
tilt angle of the Joy's law. We also showed the relationship between the ADM 
density in the solar interior (or, for example, the number of anticorrelated 
sunspots) and the Gnevyshev gap. Finally, in Section~\ref{sec-summary}, we 
provide a summary and outlook for this paper.

\section{Physics of the theory of practically empty flux tubes in strong fields of the order $\sim 10^7~G$}
\label{sec-empty-tubes}

\begin{figure*}
  \begin{center}
    \includegraphics[width=14cm]{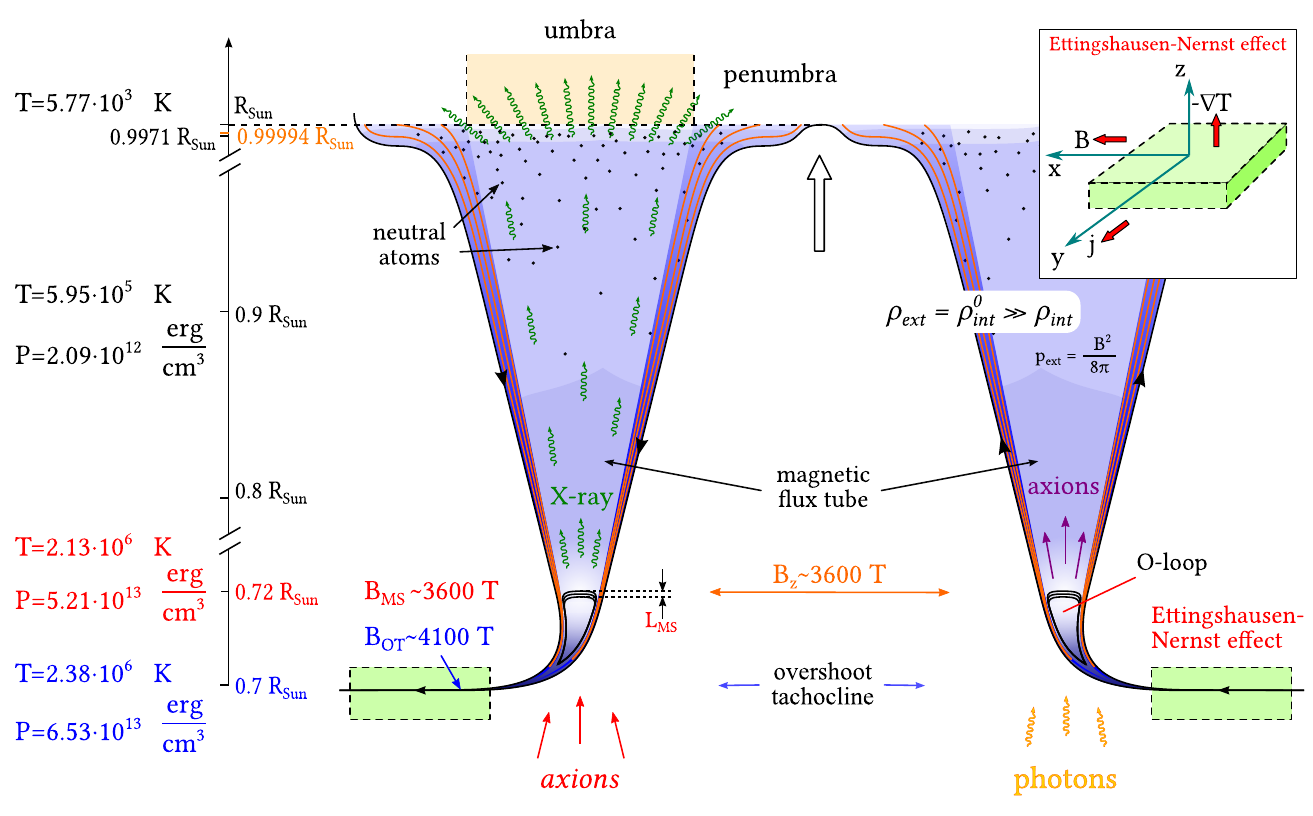}
  \end{center}
\caption{Topological effects of magnetic reconnection inside the
magnetic tubes with the ``magnetic steps'' (Fig.~B.1 in \cite{RusovDarkUniverse2021}).
The left panel shows the
temperature and pressure change along the radius of the Sun from the tachocline
to the photosphere \citep{Bahcall1992}, $L_{MS}$ is the height of the magnetic
shear steps. At $R \sim 0.72~R_{Sun}$ the vertical magnetic field,
which is developed from the horizontal part of the magnetic field (step) with
the participation of the O-loop through the well-known
Kolmogorov turbulent cascade (see Fig.~\ref{fig-Kolmogorov-cascade}),
reaches $B_z \sim 3600$~T, and the magnetic pressure $p_{ext} = B^2 / 8\pi 
\simeq 5.21 \cdot 10^{13}~erg/cm^3$ \citep{Bahcall1992}. The very cool regions 
along the entire convective zone caused by the Parker-Biermann cooling effect 
have the virtually zero internal gas pressure, i.e. the maximum magnetic 
pressure in the magnetic tubes.
The narrow ``purple'' rings between the O-loop and the tube walls 
($\rho_{ext} = \rho_{int}^0 \gg \rho_{int}$) with the Parker-Bierman cooling
effect inside, are a very important result of the existence of convective 
heating $(dQ/dt)_2$ in Section~\ref{sec-heating}.
}
\label{fig-lampochka}
\end{figure*}

Of all the
known concepts that play a significant role in understanding the connection 
between the energy transfer of a magnetic flux tube and the darkness of 
sunspots, we will consider what we believe to be the most significant. It is
based on the Parker-Biermann cooling 
effect~\citep{Parker1955a,Biermann1941,Parker1979b} in strong magnetic fields,
which explains how the result of the suppression of Parker's convective heat 
transfer manifests itself in the lower part of the magnetic flux tube (see 
Fig.~\ref{fig-lampochka}). 

\begin{figure}
\begin{center}
\includegraphics[width=8cm]{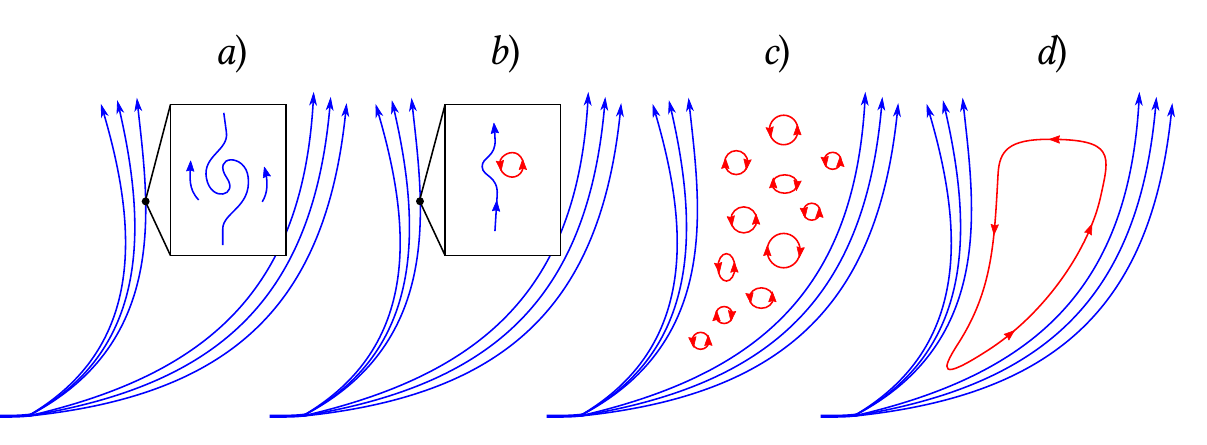}
\end{center}
\caption{Kolmogorov turbulent cascade 
\cite{Kolmogorov1941,Kolmogorov1968} and primary magnetic 
reconnection (which differs sharply from the secondary (see 
Fig.~\ref{fig-lower-reconnection}) in the lower layers inside the unipolar
magnetic tube (Fig.~B.2 in \cite{RusovDarkUniverse2021}). Common to these
various turbulent systems is the presence of the Kolmogorov's inertial region,
through which the energy is cascaded from large to small scales. In this case, 
dissipative mechanisms (as a consequence of the ``primary'' magnetic 
reconnection) overcome the turbulent energy during plasma heating.}
\label{fig-Kolmogorov-cascade}
\end{figure}

To understand the physics of the Parker's suppression of the convective heat 
transfer in strong magnetic fields, we need to tun to the dark matter axions 
originating from the Sun's core. Taking into account solar axions and the 
existence of a magnetic O-loop within the magnetic field near the tachocline 
(see Fig.~\ref{fig-Kolmogorov-cascade}d), the answer becomes very simple. 
When a magnetic O-loop forms within the magnetic field near the tachocline via
a turbulent Kolmogorov cascade (see Fig.~\ref{fig-Kolmogorov-cascade}), dark 
matter axions and high-energy photons from the radiation zone to the tachocline
undergo both axion-photon and photon-axion oscillations in this O-loop during 
the magnetic field in the tachocline, during which so-called of axion origin 
photons and photon origin axions emerge under a sunspot. This means that the 
Parker-Birman cooling effect exists due to the disappearance of barometric 
equilibrium~\citep{Parker1974a,Parker1974b} and, as a consequence, the
manifestation of the photon mean free path (Rosseland length; see Fig.~B.3 
in~\cite{RusovDarkUniverse2021}) from the tachocline to the photosphere, which 
is confirmed by the existence of both axion-photon oscillations during the 
formation of axion origin photons, and simultaneously by the existence of 
photon-axion oscillations during the formation of photon origin axions during 
MTF in the tachocline (see Fig.~\ref{fig-lampochka}).

On the other hand, high-energy photon fluxes coming from the radiation zone 
through the thin ``ring'' part of the magnetic tube, i.e. between the 
tachocline and the ``ring'' (or more precisely between the magnetic wall and 
the O-loop of the tube; see Figs.~\ref{fig-lampochka} 
and~\ref{fig-Kolmogorov-cascade}), and which is applied without the 
Parker-Biermann cooling effect, allows to determine the convective heating
$(dQ/dt)_2$ (see Sect.~\ref{sec-heating}).

\begin{figure*}
\begin{center}
\includegraphics[width=14cm]{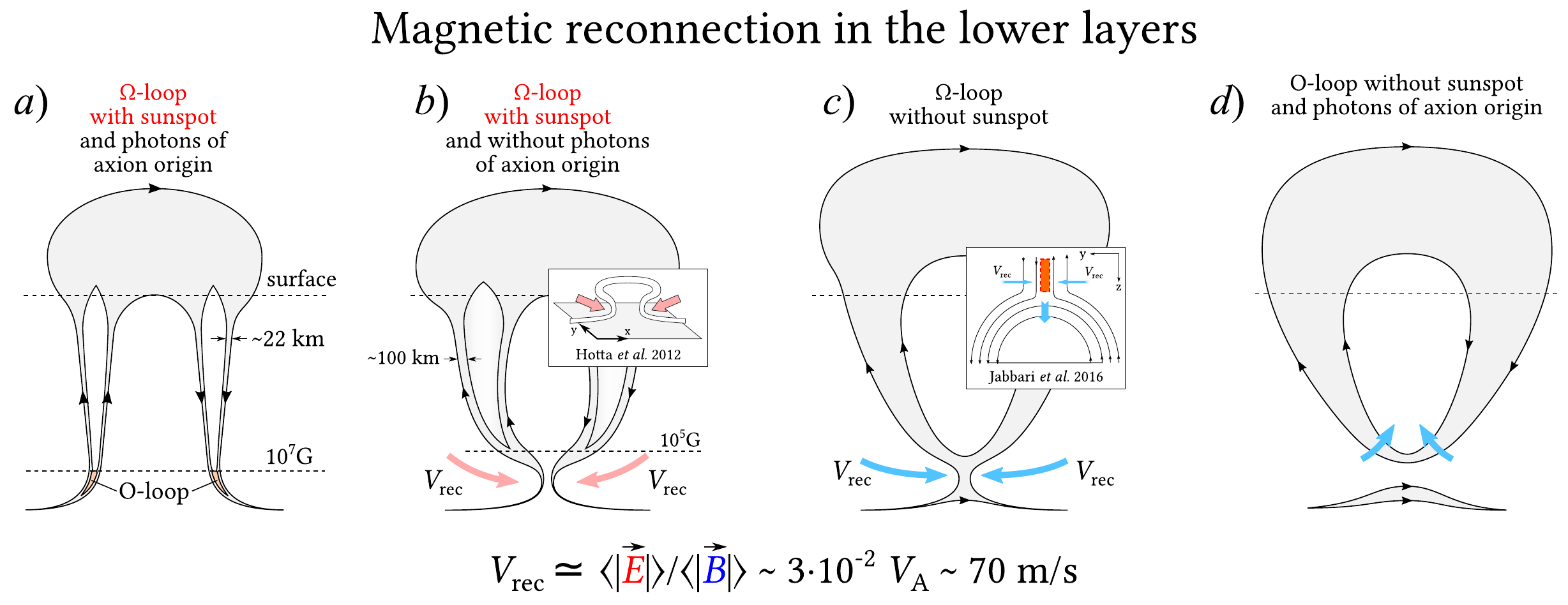}
\end{center}
\caption{Sketch of primary or secondary magnetic reconnection near the 
tachocline. (a) The $\Omega$-loop forms a sunspot shadow (with photons of axion
origin from O-loops associated with the \textbf{primary reconnection}) due to 
the indirect thermomagnetic EN effect in the tachocline, but \textbf{without 
secondary reconnection}; (b) The $\Omega$-loop with a spot in the presence of 
\textbf{secondary reconnection, but without photons of axion origin} (see b,c);
the pink arrows show the upward convective flow between the ``legs'' of the 
$\Omega$-loop as it rises from the tachocline to the visible surface of the 
Sun; (c) The $\Omega$-loop with secondary reconnection and without a sunspot;
(d) The O-loop without spots and reconnection. Passing stages (a), (b), (c) 
(from left to right), the convection around the ascending $\Omega$-loop 
``closes'' it at the base (d) and, thus, a free O-loop is formed, and the 
initial configuration of the azimuthal field is restored near the tachocline.
The blue arrows show the motion of matter leading to the connection of the 
``legs'' of the loop and their ``flying away'' from the surface of the Sun.}
\label{fig-lower-reconnection}
\end{figure*}

This solution clearly depends on the lifetime of the magnetic tubes rising from
the tachocline to the surface of the Sun. That's why due to the primary 
magnetic reconnection in the lower layers of the flux tubes (see 
Fig.~\ref{fig-Kolmogorov-cascade}d) this is not the final stage of the 
simulation. The essence of a practically empty tube in strong fields of 
$\sim 10^7~G$, which is born for the first time without a dynamo of any type
(see Fig.~\ref{fig-lower-reconnection}a), is associated with the physics of 
turbulent reconnection of magnetic bipolar structures (see 
Fig.~\ref{fig-lower-reconnection}b,c). It is very important that this is due to
the topological effects of secondary magnetic reconnection in the lower layers 
of the magnetic tube. From here, the bipolar part of the $\Omega$-loop 
rearranges itself at its base, compressing the $\Omega$-loop (blue lines; see 
Fig.~\ref{fig-lower-reconnection}b,c) and, as a consequence, organize a free 
O-loop (blue lines; Fig.~\ref{fig-lower-reconnection}d).

As described by \cite{Parker1994,Spruit1987,Wilson1990}, the upward 
convection flow around the rising $\Omega$-loop brings its ``legs'' together in
such a way that the magnetic field reconnection occurs across this loop. This 
cuts off the magnetic loop from the azimuthal magnetic field, turning it into
an O-loop (see Fig.~4 in \citep{Parker1994} and Fig.~3 in \citep{Spruit1987}). 
After that the azimuthal magnetic field restores its initial configuration and 
becomes ready for another process with $\Omega$-loop.

Let us now make some important remarks on the turbulent reconnection, the 
$\Omega$-loop transformation into the O-loop by rapid ``legs'' closure, the 
restoration of the initial azimuthal field and the preconditions for another 
$\Omega$-loop formation in the same place. It is also necessary to explain the 
physical interpretation of the overshoot process near the tachocline and 
estimate the velocity $v_{rise}$ and time $\tau_d$ of the magnetic tube rise 
from the overshoot boundary layer -- starting with the azimuthal magnetic flux 
strength of $B_{tacho} \sim 4 \cdot 10^7 ~G$ (see Eq.~A.1 in 
\citep{RusovDarkUniverse2021}).

One of the final goals of this section is to determine the general regularities
of the theory of magnetic flux tubes in strong fields of the order of 
$\sim 10^7~G$ using photons of axion origin -- the so-called universal 
Ballegooijen-Fan-Fisher model (Sec.~\ref{sec-heating}), which are generated
by the magnetic buoyancy of practically empty tubes, and as a consequence, 
their rise from the tachocline to the solar surface 
(Figs.~\ref{fig-lampochka}-\ref{fig-lower-reconnection}). Another goal is the 
essence of the physics of the primary and secondary reconnection of magnetic 
flux tubes in the lower layers of the tachocline (Sect.~\ref{sec-reconnection}), 
which is directly related to the observed features of the cycle of both double
sunspot maxima (see Sect.~\ref{sec-double-maxima}) and the tilt angle of 
Joy's law (see Sect.~\ref{sec-Joy}). Moreover, both effects (see 
Sect.~\ref{sec-heating}-\ref{sec-reconnection}) are caused by the 
existence (of dark matter?) -- solar axions generated in the core of the Sun.

\subsection{Radiative $(dQ/dt)_1$ and convective $(dQ/dt)_2$ heatings with the concern of the Parker-Biermann cooling effect and axions of photon origin}
\label{sec-heating}

The first problem is devoted to the study of the effect of virtually empty
magnetic tubes and the phenomenon of solar axions.

The assumption that the virtually empty magnetic tubes 
(Fig.~\ref{fig-lampochka}) are neutrally buoyant ($\rho_{int} = \rho_{ext}$ 
\citep{Parker1994}) implies that the temperature inside these tubes is lower 
than that of the ambient medium (Fig.~\ref{fig-lampochka} and 
Fig.~\ref{fig-lower-reconnection}a). This leads to the heat inflow, and 
consequently, the flux tube rises up (see \cite{Parker1975} or Sect.~8.8 in 
\citep{Parker1979a}). For a horizontal tube with a cross-section of radius $a$ 
the rise velocity follows from the Parker's analysis 
(\cite{Parker1975,Parker1979a}, also Eq.~(60) in \citep{Ballegooijen1982}):

\begin{equation}
v_{rise} = 2 \frac{H_p}{\tau _d} \frac{B^2}{8 \pi p_{ext}}
\left( -\delta + 0.12 \frac{B^2}{8 \pi p_{ext}} \right)^{-1},
\label{eq07-39}
\end{equation}

\noindent
where $H_p = \Re T_{ext}/g = p_{ext}/g \rho _{ext} = 0.08 R_{Sun}$ 
\citep{BohmVitense1958,Spruit1977,Brun2011} is the pressure scale height at the
tachocline, $T_{ext}$ and $p_{ext}$ are the external gas temperature and
pressure, $\delta \equiv Y = \nabla _e - \nabla _{ad} = -c_p^{-1} dS / d \xi 
= -c_p^{-1} H_p dS / dz$ is the dimensionless entropy gradient 
(see \cite{Ballegooijen1982}),
$\nabla _e \equiv d \ln T_{e} / d \ln p_{e}$ and
$\nabla _{ad} \equiv (\partial \ln T_e / \partial \ln p_e)_s$ are the local and
adiabatic temperature gradients in external and internal plasma
\citep{Spruit1974,Christensen1995,Christensen2011}, $s$ is the specific
entropy, $c_p$ is the heat capacity at constant pressure, and $\tau _d$ is the
rise time of the radiative heating of the magnetic flux tube:

\begin{equation}
\tau_d = \frac{c_p \rho a^2}{k_e} \simeq 
c_p \rho a^2
\left[ \frac{c_p  F_{tot} }{g} 
\left( 1 + \frac{2 \ell _{ov}}{5 H_p} \right)^{\nu} \right]^{-1}.
\label{eq07-40}
\end{equation}

\noindent
where for the fully ionized gas $c_p = 2.5 \Re = 2.5 p_{ext}/\rho_{ext} T_{ext}$ 
($\Re$ is the the equation of state 
$p_{ext} = \rho_{ext} \Re T_{ext} = n k_B T_{ext}$ \citep{BohmVitense1958}), 
$T(z)$ and $\rho(z)$ are the mean
temperature and density; $k_e$ is the radiative heat conductivity
(see Eq.~(35) in \cite{Ballegooijen1982});
$\ell _{ov} \approx 0.37 H_p$ \citep{Christensen1995,Christensen2011} is the
thickness of the overshoot layer; the total radiative energy flux 
$F_{tot} = L/(4 \pi r^2)$ depends on the Sun luminosity $L$; $g$ is the
gravitational acceleration.

Next we apply the condition of hydrostatic equilibrium, $dp / dz = -\rho g$,
when the adiabatic temperature gradient $(dT/dz)_{ad} = g/c_p$ may be used,
and the neutral buoyancy of the flux tube in the overshoot zone 
$(\vert \delta T \vert /T_{ext})^{-1} \sim \beta \equiv 8\pi p_{ext}/B^2$. This way we are able to
estimate the time of the radiative and/or convective diffusion $\tau_d$
(see Eq.~(\ref{eq07-40})) of the flux tube:

\begin{equation}
\tau_d = \frac{c_p \rho a^2}{k_e} \approx
\vert \delta T \vert c_p \rho 
\frac{a^2}{(1.148)^{\nu} \delta z \vert F_{tot}\vert } ,
\label{eq07-41}
\end{equation}

\noindent where

\begin{equation}
\delta z \sim (1.148)^{-\nu} \frac{1}{2} \left( \frac{a}{H_p} \right)^2
H_p \frac{\nabla _e}{\nabla _{rad}}, ~~ where ~~\nu \geqslant 3.5 ,
\label{eq07-42}
\end{equation}

\begin{equation}
F_{tot} = \frac{L}{4 \pi R_{tacho}^2} = 
H_p \frac{\nabla _{rad}}{\nabla _e} \left( \frac{dQ}{dt} \right)_1 .
\label{eq07-43}
\end{equation}

\noindent
Here $\nabla _{rad} = (\partial \ln T_{ext} / \partial \ln p_{ext})_{rad}$ is 
the radiative equilibrium temperature gradient; $(dQ/dt)_1$ is the rate of 
radiative heating, which only depends on the thermodynamic parameters $k_e$ and
$T_{ext}$ of the ambient plasma, depending only on the radial distance from the
Sun center \citep{Ballegooijen1982,Christensen1995}.

As a result, it is not difficult to show that the van Ballegooijen model
combining equations (\ref{eq07-39})-(\ref{eq07-43}) gives the final
expression for the rise time by radiation heating $(dQ/dt)_1$ from the
boundary layer of the overshoot to the solar surface,

\begin{equation}
\tau_d \approx \frac{2}{\beta} T_{ext} \left[ \frac{1}{c_p \rho_{e}} 
\left( \frac{dQ}{dt} \right)_1 \right]^{-1},
\label{eq07-44}
\end{equation}

\noindent
at which we have a very important (see section~\ref{sec-Joy}) value of 

\begin{equation}
\tau_d \approx \left. 5 \frac{1}{\beta} p_{ext} \middle/ \left( \frac{dQ}{dt} \right)_1 \right. \approx 0.81 \cdot 10^8 ~sec \approx 2 ~years ,
\label{eq08-2.6a}
\end{equation}

\noindent
by means of $1/\beta = B^2 / 8\pi \rho_{ext} \approx 2.9 \cdot 10^{-5}$, which is identical to the magnetic field $B \approx B_{eq} \sim 2 \cdot 10^5 ~G$ (see $B^2 / 8\pi = p_{ext} = nk_B T_{ext} \approx 1.665 \cdot 10^{13} ~erg \cdot cm^{-3}$ (at $0.8 R_{Sun}$ (see e.g. Fig.~\ref{fig-lampochka})) and $(dQ/dt)_1 \approx 29.7 ~erg \cdot cm^{-3} \cdot s^{-1}$ (see Eq.~(18) in~\cite{Fan1996}).

At the same time, the rate of the MFT rise from the tachocline to the solar surface
\begin{align}
v_{rise} = H_p \nabla _{ad} \frac{1}{p_{ext}} \left( \frac{dQ}{dt} \right)_1
\left( \vert \delta \vert + 0.12 \frac{B^2}{8 \pi p_{ext}} \right)^{-1}, & \nonumber \\
\nabla _{ad} = \nabla _e \simeq 0.4, &
\label{eq07-45}
\end{align}

\noindent at which we have a very important (see Sections~\ref{sec-heating}-\ref{sec-reconnection}) value of
\begin{equation}
v_{rise} \approx 1.7 \cdot 10^2 ~cm/s ,
\end{equation}

\noindent by means of $H_p \approx 5.6 \cdot 10^4 ~km$ and $(\vert \delta \vert + 0.12 B^2 / 8\pi p_{ext}) = (3/5) \times B^2 / 8\pi p_{ext} + 0.12 \times B^2 / 8\pi p_{ext} \approx 0.72 \times 2.9 \cdot 10^{-5}$ (see Eqs.~(58-60) in~\citep{Ballegooijen1982}), are almost identical to the equations (29)-(30) of~\cite{Fan1996}.

At the same time, it is easy to show that these values 
\begin{equation}
v_{rise} = \frac{2.77 H_p}{\tau_d} ,
\label{eq08-2.8}
\end{equation}

\noindent
provide a simple relationship between the rise time and the rate of ascent of
the magnetic flux tube from the overshoot boundary layer to the surface of the
Sun.

Surprisingly, these equations (Eqs.~(\ref{eq07-44}) and (\ref{eq07-45})) are 
almost identical to equations (29) and (30) from~\cite{Fan1996}, which arise
from completely different equations (60)-(61) of~\cite{Bahcall1992}.

Unlike the special van Ballegooijen model, we adopt the universal model of MFTs
with

\begin{align}
\label{eq08-2.9}
& v_{rise} = 2 \frac{H_p}{\tau _d} \frac{B^2}{8 \pi p_{ext}}
\left( -\delta + 0.12 \frac{B^2}{8 \pi p_{ext}} \right)^{-1}, \nonumber \\
& where \\
& \tau_d = \frac{c_p \rho a^2}{k_e} \left \lbrace \left( \frac{dQ}{dt} \right)_1 \left[ 1 + \left( \frac{dQ}{dt} \right)_2 \middle/ \left( \frac{dQ}{dt} \right)_1 \right] \right \rbrace ^{-1} , \nonumber
\end{align}

Here the second term $(dQ/dt)_2$ represents a convective diffusion across the
flux tube that is due to the temperature difference ($\delta T \equiv T - T_e$)
between the flux tube and the external plasma (see~\citep{Fan1996}).

Using simple calculations of equations (\ref{eq07-39}) and (\ref{eq08-2.9}) for
MFTs, it is easy to show that with the help of the
\begin{equation}
\frac{dQ}{dt} = \left( \frac{dQ}{dt} \right)_1 + \left( \frac{dQ}{dt} \right)_2
\end{equation}

\noindent
of the universal model
\begin{equation}
\tau_d \approx \frac{2}{\beta} T_{ext} \left[ \frac{1}{c_p \rho_{e}} 
\frac{dQ}{dt} \right]^{-1},
\label{eq08-2.11}
\end{equation}

\noindent and
\begin{align}
v_{rise} = H_p \nabla _{ad} \frac{1}{p_{ext}} \frac{dQ}{dt}
\left( -\delta + 0.12 \frac{B^2}{8 \pi p_{ext}} \right)^{-1},& \nonumber \\
\nabla _{ad} = \nabla _e \simeq 0.4,&
\label{eq08-2.12}
\end{align}

\noindent
is the general case of the so-called universal model of the 
van~Ballegooijen-Fan-Fisher, which is completely identical to equation (27) of
the model~\cite{Fan1996}: 
\begin{align}
\frac{d \Delta \rho}{dt} &= \rho_e \frac{v_{rise}}{H_p} \left[ \delta + \left( \frac{1}{\gamma} - \frac{2}{\gamma^2} \right) \frac{1}{\beta} - \frac{1}{\gamma}\frac{\Delta \rho}{\rho_e} \right] + \\
&\frac{2}{\gamma \beta} \rho_e \left[ \frac{\partial (v \cdot I)}{\partial s} - \vec{v} \cdot \vec{k} \right] +
\frac{\rho_e}{p_e} \nabla_{ad} \left[ \left( \frac{dQ}{dt} \right)_1 + \left( \frac{dQ}{dt} \right)_2 \right] , \nonumber
\end{align}

\noindent
where in case of quasi-static rising, it is necessary apply 
($d \Delta \rho / dt$) and $\Delta \rho$ to zero (see equation (27) 
of~\cite{Fan1996}). Furthermore, according to~\cite{Fan1996}, for simplicity,
the lifting of a uniform horizontal magnetic flux tube in the region of a 
plane-parallel overshoot region is considered, under which the conditions 
$\partial (v \cdot I) / \partial s = 0$ and $\vec{v} \cdot \vec{k} = 0$ can be
used in equation (27) of~\cite{Fan1996}: $I \equiv \partial r / \partial s$ is 
the unit vector tangential to the flux tube, and 
$k \equiv \partial ^2 r / \partial s^2$ is the tube's curvature vector.

\begin{figure}
\centering
\includegraphics[width=8cm]{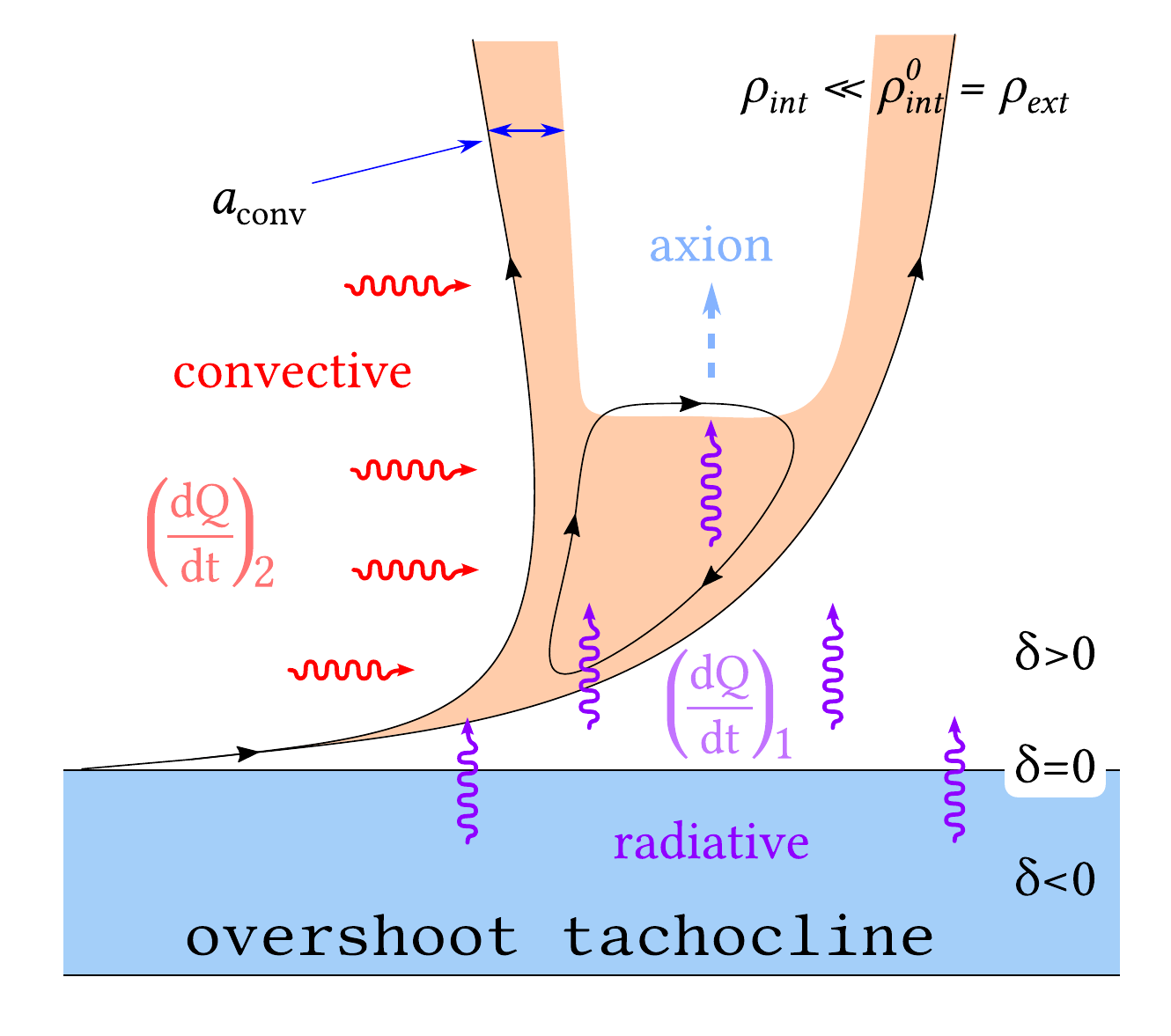}
\caption{A sketch of the magnetic tube born anchored to the
tachocline and risen to the solar surface by the neutral buoyancy
($\rho _{ext} = \rho _{int}^0$). The strong convection suppression inside the
tube leads to the abrupt decrease of temperature and density
($\rho_{int}^0 \gg \rho_{int}$), which in its turn leads to the significant
decrease in gas pressure above the umbra (sunspot). At the top of the overshoot
tachocline
the first term $(dQ/dt)_1$ characterizes the radiative heating which depends on
the thermodynamic quantities $k_{ext}$ and $T_{ext}$ of the external plasma,
changing with the distance from the center of the Sun only (see 
Eq.~(\ref{eq08-2.15})). The second term $(dQ/dt)_2$ represents the diffuse
radiation through the flux tube because of the
temperature difference between the tube ring 
(see $a \equiv a_{conv}$ between the O-loop and the tube walls) and the 
surrounding plasma (see 
Eq.~(\ref{eq08-2.16})). The keV photons (see Fig.~2 in \cite{Bailey2009}) 
coming
from the radiation zone are turned into axions in the horizontal magnetic field
of the O-loop (see Fig.~\ref{fig-lampochka}). Therefore, the radiative heating
almost vanishes in the virtually empty magnetic tube. The base of the
convection zone is defined as a radius at which the stratification switches
from almost adiabatic ($\delta = \nabla _e - \nabla _{ad} = 0$) to
sub-adiabatic ($\delta = \nabla _e - \nabla_{ad} < 0$). Meanwhile, the external
plasma turns from sub-adiabatic to super-adiabatic
($\delta = \nabla_e - \nabla _{ad} > 0$).}
\label{fig-lower-heating}
\end{figure}

It should also be recalled that, taking into account the presence of 
subadiabaticity $\delta < 0$ (see Eq.~(58) in~\citep{Ballegooijen1982}, 
Eq.~(27) in~\citep{Fan1996}, and also Fig.~\ref{fig-lower-heating}) of the
overshoot tachocline and using, according to~\cite{Ballegooijen1982}, 
non-local mixing length theory 
\citep{Ballegooijen1982,Spruit1982,BohmVitense1958,AliDey2020,Weber2015}, it can be shown that
thin, neutrally buoyant flow tubes, stable in a stratified medium, provided 
that its field strength $B$ is smaller than a critical value 
$B_c$~\citep{Ballegooijen1982}, which is approximately given by
\begin{equation}
\frac{B_c^2}{8 \pi p_{ext}} = - \gamma \delta = - \frac{5}{3} \delta .
\label{eq08-2.13}
\end{equation}

Ultimately, we call these Eqs.~(\ref{eq08-2.9})-(\ref{eq08-2.13}), in honor of
these remarkable ``solar'' physicists, the universal 
van~Ballegooijen-Fan-Fisher model (vanBFF model).

On the other hand, let us remind that on the basis of the holographic 
mechanism, generating the toroidal magnetic field in the tachocline, the 
universal model of flux tubes is predetermined by the existence of strong 
magnetic fields of the order of $B_{tacho} \sim 10^7 ~G$. Since the physics of 
the holographic mechanism does not involve a magnetic dynamo, we often refer 
to it as the universal antidynamo vanBFF model. It is determined by the 
following total energy rate per unit volume:
\begin{equation}
\frac{dQ}{dt} = \left( \frac{dQ}{dt} \right)_1 + \left( \frac{dQ}{dt} \right)_2 = 
\left( \frac{dQ}{dt} \right)_1 \left[ 1 + \frac{\alpha_1^2}{\nabla _e} \left( \frac{H_p}{a} \right)^2 \frac{1}{\beta} \right] ,
\label{eq08-2.14}
\end{equation}

\noindent where
\begin{equation}
\left( \frac{dQ}{dt} \right)_1 = - \nabla \vec{F}_{rad} = F_{tot} \frac{\nabla _e}{\nabla _{rad}} \frac{1}{H_p} = k_e \nabla_e \frac{T_e}{H_p^2} ,
\label{eq08-2.15}
\end{equation}

\begin{equation}
\left( \frac{dQ}{dt} \right)_2 = - k_e \frac{\alpha_1^2}{a^2} (T - T_e) ,
\label{eq08-2.16}
\end{equation}

\noindent
where we used the MFT heating rate which consists of the radiative heating 
rate $(dQ/dt)_1$ and convective heating rate $(dQ/dt)_2$ (see 
Eqs.~(\ref{eq08-2.14})–(\ref{eq08-2.16}), Eqs.~(13) and~(19) in \cite{Fan1996}
and Eqs.~(8)–(9) in \cite{Weber2015}); the approximate ratio 
$\vert \delta T \vert /T_{ext} \sim 1/\beta$; the parameter 
$\alpha_1^2 \approx 5.76$ \cite{Fan1996,Weber2015}; $F_{rad}$ is the radiative
energy flux (see Eq.~(15) in \cite{Fan1996}, and also Eq.~(6) and Fig.~1 in 
\cite{Weber2016}); $\nabla_e \sim 1.287 \nabla_{rad}$ (see Table~2 in 
\cite{Parker1979a}); $H_p/a$ is the factor for the lower convection zone (see 
\cite{Fan1993}), where it was previously believed that 
$a = (\Phi / \pi B_{tacho})^{1/2} \leqslant 0.1 H_p$ (see~3.2.1 in 
\cite{Ballegooijen1982}, 5.6.1 in~\cite{Fan1993}) with an average value of typical magnetic flux of 
$\Phi \sim 10^{20} - 10^{22} ~Mx$ \cite{Fan2021,Harvey1993,Sheeley1981}.

From this, we understand that it is with the strong fields that a sharp 
increase in convective heating $(dQ/dt)_2$ is necessary in the tube, which 
simultaneously leads to a sharp decrease in the outer width of the ring 
$a \equiv a_{conv}$ (between the O-loop and the tube walls in 
Fig.~\ref{fig-lower-reconnection}a) in a practically empty magnetic flux tube (see 
Fig.~\ref{fig-lampochka} and Fig.~\ref{fig-lower-reconnection}a). In this case, the 
area of the ``ring'' of the magnetic tube is approximately equal to 
$2\pi r \times a_{conv}$, where $r$ is the radius of the magnetic tube, and 
$a_{conv}$  is the width of the ``ring''.

What is the physics behind the appearance of the ``ring'' between the O-loop 
and the walls of the magnetic tube (see Fig.~\ref{fig-lower-heating}; also 
Fig.~\ref{fig-lampochka})? In simple words, one can say the following. The 
appearance of the ``ring'' cross-section is, as a consequence, the result of 
the production of both axion-origin photons (by converting solar axions into 
photons in the tachocline) and photon-origin axions (through the conversion of 
high-energy photons from the radiation zone to axions in the tachocline). From 
here, on the one hand, the axions of photonic origin are the part of the 
manifestation of the mean free path of axion origin photons from the tachocline
to the photosphere. On the other hand, they are the part of the manifestation 
of magnetic tube ``ring'', where the convective heating $(dQ/dt)_2$ dominates 
over the radiation heating $(dQ/dt)_1$.

At the same time, we remember that anticorrelated 11-year ADM density 
modulations in the Sun interior and the ADM
around BH (see Sect.~3 in~\citep{RusovDarkUniverse2021}) are not only 
interconnected, but simultaneously, as consequence, are driving the 11-year 
modulations of the solar axion density (see Eqs.~(27)-(28) in 
\citep{RusovDarkUniverse2021}).

As a result, assuming 11-year variations in convective heating $(dQ/dt)_2$ in 
the form of a maximum and minimum in the width of the ``ring'' 
$a \equiv a_{conv}$ (between the O-loop and the tube walls in 
Fig.~\ref{fig-lower-heating}), we apply a new analysis of the universal vanBFF
model (see Eqs.~(\ref{eq08-2.14})-(\ref{eq08-2.16}) similar to the last part of
Eqs.~(13) and~(19) in~\citep{Fan1996})), where the values such as the magnetic
flux $\Phi$ (see~\cite{Harvey1993,Sheeley1981}) and the magnetic fields at the
tachocline ($3.6 \cdot 10^7 ~G$ in Fig.~\ref{fig-lampochka}) are consistent 
with known observational data:

%
%
\noindent

%
\textbf{$\bullet$} values of the magnetic tube (see 
Fig.~\ref{fig-lower-reconnection}a), based on the physics of convective heating
$(dQ/dt)_2$, are related to the radius of the ``ring'' between the O-shaped 
loop and the walls of the magnetic tube (Fig.~\ref{fig-lower-heating}), 
and correspond exactly to the ``observed'' values of the ratio between the 
maximum and minimum magnetic flux and the magnetic field at the 
tachocline~\cite{Harvey1993,Sheeley1981,Fan1993}. Therefore, we are interested 
in the observed magnetic fluxes
\begin{equation}
\Phi _{max} = 2 \pi r (a_{conv})_{min} (B_{tacho})_{max} \approx 10^{22} ~Mx ,
\label{eq-08-22-Phimax}
\end{equation}

\noindent
where $\Phi _{max} \approx 10^{22} ~Mx$  is at the maximum of the ``measured''
magnetic fluxes~\cite{Harvey1993,Sheeley1981} and the maximum of the 
``measured'' magnetic field MFT $(B_{tacho})_{max} \sim 3.7\cdot 10^7 ~G$ 
(see Fig.~\ref{fig-lampochka}), which is associated with the so-called 
Ettingshausen-Nernst thermomagnetic effect, caused by the Parker-Biermann 
cooling effect.


Then a minimum of magnetic flux appears
\begin{equation}
\Phi _{min} = 2 \pi r (a_{conv})_{max} (B_{tacho})_{min} \approx 5 \cdot 10^{21} ~Mx ,
\label{eq-08-23-Phimin}
\end{equation}

\noindent
where $\Phi _{min} \approx 5 \cdot 10^{21} ~Mx$ is at the minimum of the 
``measured'' magnetic fluxes~\cite{Harvey1993,Sheeley1981} and the minimum of 
the ``measured'' magnetic field MFT $(B_{tacho})_{min} \sim 3.7\cdot 10^6 ~G$.



It is very important here that, on the one hand, the anchored magnetic field in
the tachocline rises from the bottom of the convective zone to the solar
photosphere in the form of thin isolated filaments known as flux tubes (see e.g.
\citep{Parker1979a}), and on the other hand, the well-known nonlinear dynamics
of a thin magnetic tube was modeled in the form that has been used by 
\cite{Spruit1977}. In this case, the magnetic tube is assumed to be thin in the
sense that its cross-sectional radius $a_{conv}$ is negligibly small with 
respect to both the atmospheric scale height (i.e. $a_{conv} \ll H_p$ (see 
Fig.~\ref{fig-lower-heating})) and any variation scales along the tube. This 
means that each magnetic field line is roughly parallel to the axis of the 
tube, and the Spruit's model is applicable only to untwisted magnetic tubes.

This raises a rather complicated but simple question: In what way, without 
twisting, can a real magnetic tube, which is not integral and will not behave
as a single object for a very long time 
\citep{Parker1979a,Spruit1977}, be obtained without destruction by
hydrodynamic forces?

The answer, oddly enough, is very simple! For an untwisted thin magnetic tube,
the pressure balance is always maintained not only along the entire radius of
the ``ring'' between the O-shaped contour and the walls of the magnetic tube 
(Fig.~\ref{fig-lower-heating}), which is solved without destruction by 
hydrodynamic forces, but at the same time leads to the suppression of 
convection along the inner radius of the O-shaped contour due to a sharp 
decrease in temperature and density (see $\rho_{int}^0 \gg \rho_{int}$ in 
Fig.~\ref{fig-lower-heating}). The latter means that in very strong magnetic
fields (e.g. $5 \cdot 10^5 ~G < B < 10^7 ~G$) an untwisted thin magnetic tube
will, surprisingly, always be perpendicular to the azimuthal direction from the
tachocline to the surface of the photosphere!

The physics of this process is briefly as follows. The high-energy photons 
from the radiation zone through axion-photon oscillations in the O-loop inside
the magnetic tube near the tachocline create the so-called axions of photon 
origin under the sunspot. This means that at such strong magnetic fields, the 
Parker-Biermann effect of cooling inside the magnetic tube, where the Sun's 
gravitational field pulls gas out of the inner tube (as described by 
hydrostatic pressure, known as the barometric law $dp/dz = -\rho g$), develops 
due to the ``disappearance'' of the  convective heat transfer $(dQ/dt)_2$ and, 
as a consequence, the temperature in the lower part of the magnetic tube with 
the help of axions of photon origin from photon-axion oscillations in the 
O-loop near the tachocline. As a result, it opens a free path (in the radial 
direction!) for photons of axion origin (Rosseland length; see Fig.~B.3 in 
\citep{RusovDarkUniverse2021}) from the tachocline to the photosphere (see 
Fig.~\ref{fig-lower-reconnection}a)!

From here, using Eq.~(\ref{eq-08-22-Phimax}) and Eq.~(\ref{eq-08-23-Phimin}), 
we obtain the cross-section values $a_{conv}$ (between the O-loop and the tube
walls in Fig.~\ref{fig-lower-reconnection}a) at the minimum 
\begin{equation}
(a_{conv})_{min} \sim 22 ~km \approx 3.7 \cdot 10^{-4} H_p
\label{eq-08-24-aconvmin}
\end{equation}

\noindent and at the maximum
\begin{equation}
(a_{conv})_{max} \sim 220 ~km \approx 3.7 \cdot 10^{-3} H_p ,
\end{equation}

\noindent
at which the rise time of convective heating of the flux tube from the 
tachocline to the photosphere (see Eq.~(\ref{eq08-2.11})) corresponds to the
equations


\begin{align}
&(\tau_d)_{conv} = \frac{2}{\beta} T_{ext} \left \lbrace \frac{1}{c_p \rho_{ext}} 
\left( \frac{dQ}{dt} \right)_1
\left[ 1 + \frac{\alpha _1 ^2}{\nabla_e} \frac{1}{\beta}
\left( \frac{H_p}{a_{conv}} \right)^2 
\right] \right \rbrace ^{-1}    \label{eq08-2.19} \\
& = \frac{2}{\beta} T_{ext} 2.5 \Re \rho_{ext} \left \lbrace 
\left( \frac{dQ}{dt} \right)_1
\left[ 1 + \frac{\alpha _1 ^2}{\nabla_e} \frac{1}{\beta}
\left( \frac{H_p}{a_{conv}} \right)^2 
\right] \right \rbrace ^{-1} \nonumber \\
& = \frac{5}{\beta} T_{ext} (p_{ext}/ \rho_{ext} T_{ext}) \rho_{ext} \left \lbrace 
\left( \frac{dQ}{dt} \right)_1
\left[ 1 + \frac{\alpha _1 ^2}{\nabla_e} \frac{1}{\beta}
\left( \frac{H_p}{a_{conv}} \right)^2 
\right] \right \rbrace ^{-1} \nonumber \\
& = \frac{5}{\beta} p_{ext}\left \lbrace 
\left( \frac{dQ}{dt} \right)_1
\left[ 1 + \frac{\alpha _1 ^2}{\nabla_e} \frac{1}{\beta}
\left( \frac{H_p}{a_{conv}} \right)^2 
\right] \right \rbrace ^{-1}  \label{eq08-2.19a}  \\
& = \frac{5}{\beta} nk_B T_{ext} \left \lbrace 
\left( \frac{dQ}{dt} \right)_1
\left[ 1 + \frac{\alpha _1 ^2}{\nabla_e} \frac{1}{\beta}
\left( \frac{H_p}{a_{conv}} \right)^2 
\right] \right \rbrace ^{-1}  \label{eq08-2.19b} \\
& \cong \frac{5}{\beta} \rho_{ext} \mu^{-1} m_p^{-1} k_B T_{ext} \left \lbrace 
\left( \frac{dQ}{dt} \right)_1
\left[ 1 + \frac{\alpha _1 ^2}{\nabla_e} \frac{1}{\beta}
\left( \frac{H_p}{a_{conv}} \right)^2 
\right] \right \rbrace ^{-1} \label{eq08-2.19c} 
\end{align}

\noindent
where at the maximum the values of 
$(a_{conv})_{min} \approx 3.7\cdot 10^{-4} H_p$ and
$1/\beta = B^2 / 8 \pi p_{ext} \approx 1$ we represent 11-year variations of 
the magnetic flux tube in the photosphere for sunspots (see (\ref{eq08-2.19b})
or (\ref{eq08-2.19a}), in the form

\begin{align}
&(\tau_d) _{conv}^{max} \approx 5 \times (6.5 \cdot 10^{13} ~erg \cdot cm^{-3}) / 3 \cdot 10^9 ~erg \cdot cm^{-3} \cdot s^{-1} \nonumber \\
&\approx 10^{15} ~erg \cdot cm^{-3} / 3 \cdot 10^9 ~erg \cdot cm^{-3} \cdot s^{-1} \approx 1.1 \cdot 10^5 ~s ,
\end{align}

\noindent or

\begin{equation}
(\tau_d) _{conv}^{max} \approx 1.1 \cdot 10^5 ~s \sim 1.3 ~day , 
\label{eq08-2.19a2}
\end{equation}

\noindent
where $p_{ext} = 6.53 \cdot 10^{13} ~erg / cm^3$ (see \citep{Bahcall1992}), 
$T_{ext} = 2.3\cdot 10^6 ~K$ and $\rho_{ext} = 0.2 ~g/cm^3$ 
\citep{Bahcall1992}; $c_p = 2.5 \Re = 2.5 p_{ext} / \rho_{ext} T_{ext}$ 
($\Re$ is the gas constant in the equation of state 
$p_{ext} = \rho_{ext} \Re T_{ext} = nk_B T_{ext}$ \citep{BohmVitense1958}); 
$\mu = 0.58$ is mean molecular weight of hydrogen in overshoot tachocline, 
$m_p \cong  1.67 \cdot 10^{-24} ~g$ is the proton mass: "2018 CODATA Value: 
proton mass", The NIST Reference on Constants, Units, and Uncertainty, 20~May 
2019; 
$k_B = 1.38 \cdot 10^{-23} ~J/K$; $(dQ/dt)_1 \approx 29.7 ~erg \cdot cm^{-3} 
\cdot s^{-1}$ (see Eq.~(18) in \citep{Fan1996}) is the rate of radiative 
heating; $\alpha_1^2 \approx 5.76$ \citep{Fan1996,Weber2015}; 
$\nabla_e \approx 0.4$ (see Table~2 in \citep{Spruit1974}); 
$H_p = \Re T_{ext}/g = p_{ext} / g \rho_{ext} = 0.08 R_{Sun}$ 
\citep{BohmVitense1958,Spruit1977,Brun2011}; 
$a_{conv}^{min} \sim 3.7 \cdot 10^{-4} H_p$ (see 
Eq.~(\ref{eq-08-24-aconvmin})).

%
%

And, as a consequence, the magnitude of the MFT rise speed (see 
Eq.~(\ref{eq08-2.12})) will be equal to
\begin{equation}
(v_{rise})_{conv} = \frac{2.77 H_p}{(\tau_d)_{conv}}  ,
\label{eq-08-31}
\end{equation}

\noindent
both at the maximum of the solar cycle
\begin{equation}
(v_{rise})_{conv}^{max} \sim 1.40 ~km/s   ,
\label{eq08-2.21b}
\end{equation}

\noindent
which are almost identical to the observational data of the known works (see
about 1.4~km/s 
in~\cite{Ilonidis2012,Ilonidis2013,Kosovichev2016,Kosovichev2018}), and at the
minimum of the solar cycle
\begin{equation}
(v_{rise})_{conv}^{min} \sim 0.14 ~km/s   ,
\label{eq08-2.21c}
\end{equation}

\noindent
which are almost identical to the observational data of the known works (see 
about $\sim 0.15 ~km/s$ in~\cite{Birch2016}), from which we
obtain the ``minimum'' time of increase of convective heating $(dQ/dt)_2$ of
the flux tube from the tachocline to the photosphere (see 
Eq.~(\ref{eq08-2.11})), which corresponds to the equation
\begin{equation}
(\tau_d) _{conv}^{min} \approx 10^6 ~s \sim 10 ~days ,
\label{eq08-2.20c}
\end{equation}

\noindent
where $p_{ext} = (nk_B T)_{min} = 3.56 \cdot 10^{13} ~erg/cm^3$, 
$T_{ext} = 10^6 ~K$, $\rho_{ext} \approx 0.25 ~g/cm^3$ (see 
Eq.~(\ref{eq08-2.19a})). 

Ultimately, these Eq.~(\ref{eq-08-31}) are the source of the first (fast in 
time) maximum of solar activity (e.g. sunspots) from the observed double 
maxima of 11-year cycles, while the second (two-year in time) maximum of 
sunspots is associated with observational data on the slope of Joy's law 
(see Fig.~\ref{fig-prim-sec-reconnection}b$_2$,c on the right).

This raises an intriguing point regarding helioseismology. One of the first 
solutions in our section is the buoyancy of a virtually empty magnetic flux 
tube in strong magnetic fields of $\sim 10^7 ~G$ near the tachocline, 
determined by the thermomagnetic Ettingshausen-Nernst effect (see Apendix~A 
in~\cite{RusovDarkUniverse2021}), manifesting itself as ``visible'' sunspots on
the solar surface. This is due to the fact that helioseismological inversions, 
which are well suited not only for studying the rotation and structure of the 
Sun, can sometimes be well suited due to the mediated effects of very strong 
magnetic fields at empty magnetic flux tubes. Let us repeat once again that it 
is the existence of empty magnetic flux tubes that allows photons of axion 
origin to ``rush'' past the tachocline to the photosphere both at high speed 
(see $\sim 1.4 ~km/s$ 
in~\cite{Ilonidis2012,Ilonidis2013,Kosovichev2016,Kosovichev2018}) and at short
times (see $\sim 1.3 ~days$ for Eqs.~(\ref{eq08-2.19a2})-(\ref{eq-08-31}))! 
Thus, they can unequivocally confirm the presence of a magnetic field in the 
solar tachocline exceeding $>10^5 ~G$ to, for example, $10^7 ~G$. On the other
hand, it is the existence of ``visible'' sunspots in such strong fields that 
causes the dynamo effect to completely disappear.

The answer, surprisingly, is quite simple. It means that the ``measured'' 
velocity (maximum about 
$\sim 1.4 ~km/s$~\cite{Ilonidis2012,Ilonidis2013,Kosovichev2016,Kosovichev2018}
and minimum about $\sim 0.15 ~km/s$ in~\cite{Birch2016}) of magnetic flux 
tubes, which are associated with structural changes, for example, in the form 
of cross-sections of these tubes at the maximum ($\sim 10^{22} ~Mx$) or minimum
($5\cdot 10^{21} ~Mx$) of the ``measured'' magnetic 
fluxes~\cite{Harvey1993,Sheeley1981}, can be easily determined at the maximum 
$\sim 10^7 ~G$ or minimum $\sim 10^6 ~G$ values of the magnetic fields (see 
Eqs.~(\ref{eq-08-22-Phimax})-(\ref{eq-08-23-Phimin})). The modernity of this 
considered helio-seismological inversion remains to this day!

In other words, the modernity of this well-thought-out helioseismological 
inversion remains to this day!

\subsection{Physics of primary and secondary reconnection of magnetic flux tubes in the lower layers of the tachocline}
\label{sec-reconnection}

The problem is devoted to the physics of primary and secondary magnetic 
reconnection in the lower layers of the tube, which is directly related to the
observed features of the cycle of both double maxima of sunspots and the tilt 
angle of Joy's law.

First, let us recall the essence of the primary reconnection inside magnetic 
tubes near the tachocline ($B \sim 10^7 ~G$), which generates the O-loop (see 
Fig.~\ref{fig-lower-reconnection}d), and thus participates in the formation of
axion origin photons (see Fig.~\ref{fig-lampochka}, left) and axions of photon 
origin (see Fig.~\ref{fig-lampochka}, right) from the O-loop to the 
photosphere, is the source of sunspots, which (due to the cooling of the 
Parker-Biermann effect inside the tube) rise from the tachocline to the surface
of the Sun.

\begin{figure*}
\begin{center}
\includegraphics[width=14cm]{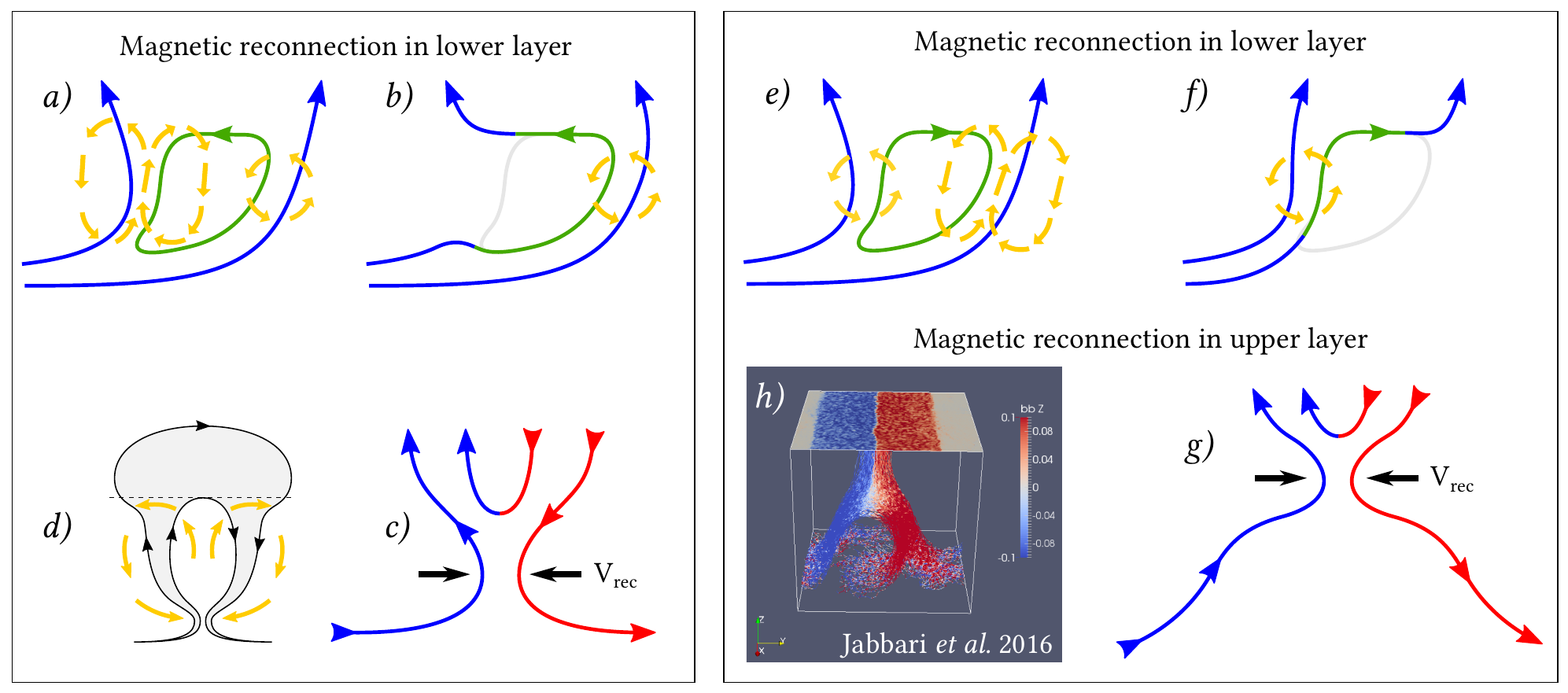}
\end{center}
\caption{Topological effects of secondary magnetic reconnection in the lower (left) or upper (right) layers of the magnetic tube. Here, the unipolar part of the $\Omega$-loop is rearranged at its base, compressing the $\Omega$-loop (blue lines) and, as a consequence, organizing a free O-loop (green lines) (\textit{a},\textit{e}) to form (blue-yellow lines) the first part of the secondary magnetic reconnection (\textit{b},\textit{f}).
The yellow lines show the movement of matter leading to the loop connection ``leg'' (see analogous Fig.~4 in \cite{Parker1994}). If the O-loop (primary magnetic reconnection, green lines) can randomly have different directions of magnetic fields, then the secondary magnetic reconnection can generate loop ``legs'' in different layers, for example, in lower layers (\textit{c} and \textit{d} as an analog of Fig.~\ref{fig-lower-reconnection}b) and upper layers (\textit{e} and \textit{f} as an analog of Fig.~4 in \cite{JabbariEtAl2016}).}
\label{fig-upper-reconnection}
\end{figure*}

On the other hand, we are interested in the essence of the very important 
secondary magnetic reconnection (see Fig.~\ref{fig-upper-reconnection}b), which
is the source of the observed double maxima of one sunspot cycle -- it shifts 
by 2.5~years between the primary and secondary solar maxima when the magnetic 
tube rises, and is associated with both the observational data on the tilt 
angle of Joy's law and the ``disappearance'' of sunspots on the solar surface.
 
In Fig.~\ref{fig-upper-reconnection} one can see the conditions for the 
secondary reconnection between the O-loop (green lines) and the unipolar part 
of the $\Omega$-loop (blue lines) that can organize them in the lower layer 
(Fig.~\ref{fig-upper-reconnection}, left) or in the upper layer 
(Fig.~\ref{fig-upper-reconnection}, right), thereby showing the appearance of
bipolar magnetic tubes in various versions 
(Fig.~\ref{fig-upper-reconnection}c,d and g,h).

From here, since the sunspot cycle is identical to the maxima and minima of the
magnetic tube number on the surface of the Sun, we are interested in the 
physics of the maxima and minima both during the primary reconnection for the 
time $(\tau_d)_{conv}$ of the rise of the almost empty magnetic tube from the 
tachocline to the solar photosphere (see about $\sim 1-2$ days at max in 
Eq.~(\ref{eq08-2.19a2}) and about $\sim 10$ days at min in 
Eq.~\ref{eq08-2.20c}), and during the secondary reconnection $(\tau_n)_{rec}$,
which are associated with the features of the tilt angle of Joy's law (see 
Fig.~\ref{fig-prim-sec-reconnection}c), during the rise of the almost empty 
magnetic tube from the surface of the Sun (see about $\sim 2$ years in 
Fig.~\ref{fig-prim-sec-reconnection}b$_2$ and c) and, as a consequence, their 
``disappearance'' on the Sun (see about $<0.5$ years in 
Fig.~\ref{fig-prim-sec-reconnection}c).

\subsubsection{Secondary reconnection of the MFT without axions, but with concern of turbulent diffusion and the tilt angle of the Joy law}
\label{sec-Joy}

The first part of the scenario consists in discussing the physics of magnetic 
flux tube buoyancy at strong fields and with the participation of axion origin
photons (see Fig.~\ref{fig-axion-constraints}), associating with the formation
of an O-loop (see the existence of primary magnetic reconnection in 
Fig.~\ref{fig-lampochka}) inside the magnetic flux tube near the tachocline.

\begin{figure}
\begin{center}
\includegraphics[width=\linewidth]{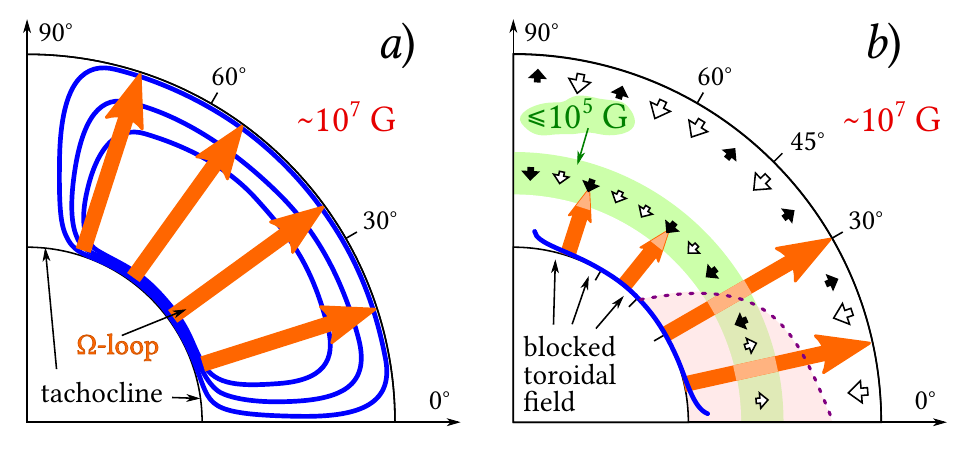}
\end{center}
\caption{Scheme of turbulent reconstruction of the toroidal magnetic field in
the convective zone (CZ): \textbf{(a)} meridional circulation (closed blue 
poloidal field lines), which is the source of the toroidal field in the 
tachocline (thin black lines), and the magnetic buoyancy of the 
flux tubes (red arrows) with a magnetic field of 
$\sim 10^7 ~G$;
\textbf{(b)} In the CZ, through secondary turbulent reconnection of the 
magnetic tube near the tachocline (see 
Fig.~\ref{fig-prim-sec-reconnection}b$_2$,c), effects of both macroscopic 
turbulent plasma diamagnetism are observed, facilitating the formation of two 
layers on the surface (see analogous 
\citep{Krivodubskij2005,Krivodubskij2021,Krivodubskii1992}) with different 
vertical directions of horizontal field transfer (see ``black'' arrows), and 
the conditions of magnetic field transfer along $\nabla \rho$ due to 
small-scale magnetic pulsations, acting against buoyancy. However, in the deep
layers in the near-equatorial region, the $\nabla \rho$ effect provides upward
advection, which, on the contrary, directs magnetic buoyancy, and thus, can 
facilitate the penetration of strong fields to the surface.}
\label{fig-meridional-cut}
\end{figure}

For the condition of neutral buoyancy
($\rho_{int} = \rho_{ext} \approx 0.2 ~g/cm^3$~\cite{Bahcall1992}; see A.16 
in~\cite{RusovDarkUniverse2021}, Figs.~\ref{fig-lampochka} 
and~\ref{fig-lower-heating}) and the strong toroidal field of the magnetic 
tube, $B_{Sun} = 4.1 \cdot 10^7 ~G$, as well as the condition of the average 
width of the ``thin'' ring $a_{conv}^{min} \sim 3.7 \cdot 10^{-4} ~H_p$ (see 
Eq.~(\ref{eq-08-24-aconvmin})) between the O-shaped loop and
the walls of the magnetic tube (see Fig.~\ref{fig-lower-heating}), it is easy 
to recall the estimate of the time $(\tau_d)_{conv}^{max} \sim 1.3 ~days$ (see 
Eq.~(\ref{eq08-2.19a2})) and velocity $(v_{rise})_{conv}^{max} \sim 1.40 ~km/s$
(see Eq.~(\ref{eq08-2.21b}), as well as the time 
$(\tau_d)_{conv}^{min} \sim 10 ~days$ (see Eq.~(\ref{eq08-2.20c})) and velocity $(v_{rise})_{conv}^{min} \sim 0.14 ~km/s$ (see Eq.~(\ref{eq08-2.21c}) of the convective rise from the 
tachocline to the photosphere, which for such large magnetic fields has a 
significant amount of rise of the magnetic tubes at all latitudes (see 
Fig.~\ref{fig-lower-reconnection}a and the red arrows of the magnetic flux 
tubes in Fig.~\ref{fig-meridional-cut}a).

This solution clearly depends on the rise time of the magnetic tubes rising 
from the tachocline to the solar surface. Therefore, the primary magnetic 
reconnection itself in the lower layers of the magnetic tubes (see 
Fig.~\ref{fig-lower-reconnection}a), is not the final stage of the modeling. 
The essence of a practically empty tube in strong fields of $\sim 10^7 ~G$ is
related (through the primary magnetic reconnection) to the physics of the 
secondary turbulent reconnection of magnetic bipolar structures (see 
Fig.~\ref{fig-lower-reconnection}b, and also Fig.~4 in \citep{Brun2011}). 
Here, on its basis, the bipolar part of the $\Omega$-loop is reconstructed, 
compressing the $\Omega$-loop (blue lines: see 
Fig.~\ref{fig-lower-reconnection}b,c) near the tachocline and simultaneously 
lifting them (from 2 to 3 years) to the surface of the Sun, and, as a 
consequence, after some time (from several days to several months (see 
\cite{Petrovay1997,Kichatinov1991})
 a free O-loop appears (blue lines in 
Fig.~\ref{fig-lower-reconnection}d), which eventually ``disintegrates'' or, 
more precisely, the O-loop (without spots) flies away from the surface of the
Sun (see Fig.~\ref{fig-lower-reconnection}d).

This raises a rather unexpected but important question: How is the secondary 
magnetic reconnection of flux tubes with 
$B_{tacho}^{Sun} = B_z = 4.1 \cdot 10^7 ~G$, leading to a real sharp decrease
in the magnetic field to $B_{eq} \leqslant 2 \cdot 10^5 ~G$ at 
$\sim 0.80 R_{Sun}$ (see e.g. Fig.~\ref{fig-prim-sec-reconnection}b,c), related
to the existence of downward turbulent diamagnetic pumping and rotational 
$\nabla \rho$-pumping at relatively low magnetic fields, and as a consequence, 
at low escape velocities of flux tubes from the solar surface?

In this regard, we obtain an answer to this important question. For this 
purpose, we consider the second part of the scenario, based on the remarkable
idea of L.L.~Kitchatinov~\cite{Kichatinov1991}, which includes the generation
of the magnetic field of $\leqslant 2\cdot 10^5 ~G$ near the bottom of the 
convective zone and transfer of the toroidal field from the deep layers at 
different latitudes. It is very important that the efficiency of the magnetic 
buoyancy transfer is associated with the participation of two processes: 
macroscopic turbulent 
diamagnetism~\cite{Zeldovich1957,Zeldovich1956,Spitzer1957} and rotational
$\nabla \rho$-pumping \citep{Kichatinov1991}.

Let us note some important properties of macroscopic turbulent diamagnetism. It
is known that Zel'dovich~\cite{Zeldovich1956,Zeldovich1957} and 
Spitzer~\cite{Spitzer1957} in 1957 discovered the diamagnetism of 
inhomogeneously turbulent conducting liquids, in which the inhomogeneous 
magnetic field moves as a single whole. In this case the turbulent fluid, for
example, with nonuniform effective diffusivity $\eta _T \approx (1/3) v l$ (see
Fig.~1 in~\cite{Kitchatinov2008}, and 
also~\cite{Zeldovich1957,Zeldovich1956,Spitzer1957}) behaves like a diamagnetic
one and carries the magnetic field with the effective 
velocity~\cite{Kohler1973}
\begin{equation}
\vec{v}_{dia} = - \frac{1}{2} \nabla \eta_T ,
\label{eq07-68}
\end{equation}

\noindent
where $l$ is the mixing length of turbulent pulsations, and 
$v = \sqrt{\langle v^2 \rangle}$ is the root-mean-square velocity of turbulent 
motion. The minus sign on the right in Eq.~(\ref{eq07-68}) shows the meaning of
turbulent magnetism: it is not paramagnetic magnetism, so magnetic fields repel
from regions with relatively high turbulent intensity. In other words, 
macroscopic turbulent plasma diamagnetism and, as a consequence, the so-called 
macroscopic diamagnetic effect (see \cite{Radler1968a,Radler1968b}) in the 
physical sense is the displacement of the averaged magnetic field $B$ from 
regions with increased intensity of turbulent pulsations to regions with less 
developed turbulence~\cite{Krause1980,Kichatinov1992}.

However, there is an interesting problem of the diamagnetic process caused by 
inhomogeneous turbulent intensity with allowance for the total nonlinearities 
in the magnetic field. This is due to the fact that up to the present time 
analytical estimates have been obtained only for the limiting cases of weak and
strong magnetic fields. For example, at strong magnetic fields of flux tubes, 
at $B > 10^5 ~G$, e.g. $\geqslant 10^7 ~G$, the diamagnetic effect becomes 
almost negligible, in particular, strong magnetic damping of diamagnetism 
$\sim B^{-3}$ is obtained for super-equipartitions of fields 
\citep{ChenF2022} when turbulence is close to two-dimensional 
\citep{Zeldovich1957}. On the other hand, for very weak fields the diamagnetic 
pumping, which is predetermined by the intensity of turbulence at 
$B < 10^5 ~G$, is a very effective process 
\citep{ChenF2022}.

Among the known limiting cases of weak and strong magnetic fields from the 
tachocline to the surface, we are interested, in the addition to the portions
of strong fields (via the secondary Lazaryan-Vishniac reconnection (see ~\cite{Lazarian1999,Lazarian2004,Kowal2009}, and also §III and §IX 
in~\cite{Lazarian2020}) 
near the tachocline, which have a very slow Sweet-Parker velocity $(V_{rec})_2$
(see~\cite{Parker1957,Sweet1958}) for 
Fig.~\ref{fig-prim-sec-reconnection}b$_2$), the part of the weak magnetic field
from $0.8~R_{Sun}$ to near $0.85~R_{Sun}$ (see 
Fig.~\ref{fig-prim-sec-reconnection}b$_2$,c), which constitutes the 
super-equipartitions of field $B \geqslant B_{eq}$ and is the source of two 
processes: diamagnetic pumping, and rotational $\nabla \rho$-pumping.
This is due to the fact that the secondary magnetic reconnection of the flux 
tubes (see Figs.~\ref{fig-upper-reconnection}b,d,f,g) leads to the real 
increase in the upward magnetic field to $B \approx 2 \cdot 10^5 ~G$ at 
$\sim 0.8~R_{Sun}$, which is identical to the 
$1/\beta = B^2 / 8\pi p_{ext} \approx 2.9 \cdot 10^{-5}$ (see e.g. 
Eq.~(\ref{eq08-2.6a}), and Fig.~\ref{fig-prim-sec-reconnection}b$_2$,c). 
This means that the real toroidal magnetic flux tube is a consequence of the 
formation of the secondary reconnection, which generates both the 
``anti-buoyant'' effects of downward turbulent diamagnetic transport
and the rotational effect of the magnetic $\nabla \rho$-pumping at polar 
latitudes, and the ``buoyant'' effects of magnetic flux tubes at equatorial 
latitudes as a result of the upward rotational effect of magnetic 
$\nabla \rho$-pumping.

\begin{figure*}
\begin{center}
\includegraphics[width=\linewidth]{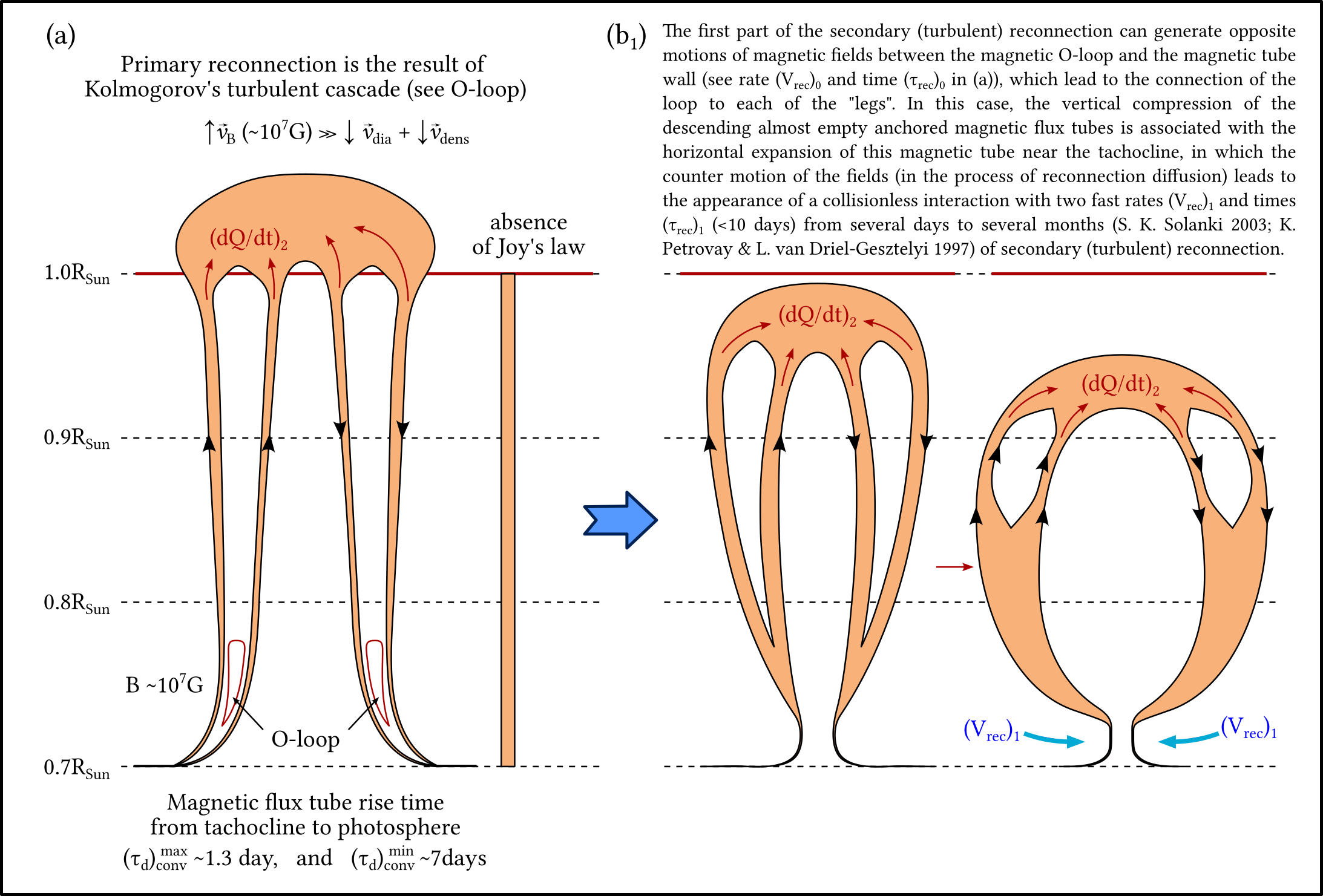}
\end{center}
\caption{An alternative theory of magnetic flux tubes in strong fields by 
means of axion origin photons associated with primary and secondary 
reconnections. [Ref: S.K.~Solanki~(2003)~\cite{Solanki2003}, 
K.~Petrovay \& L.~van~Driel-Gesztelyi~(1997)~\cite{Petrovay1997}.]}
\label{fig-prim-sec-reconnection}
\end{figure*}
\begin{figure*}
\ContinuedFloat
\begin{center}
\includegraphics[width=\linewidth]{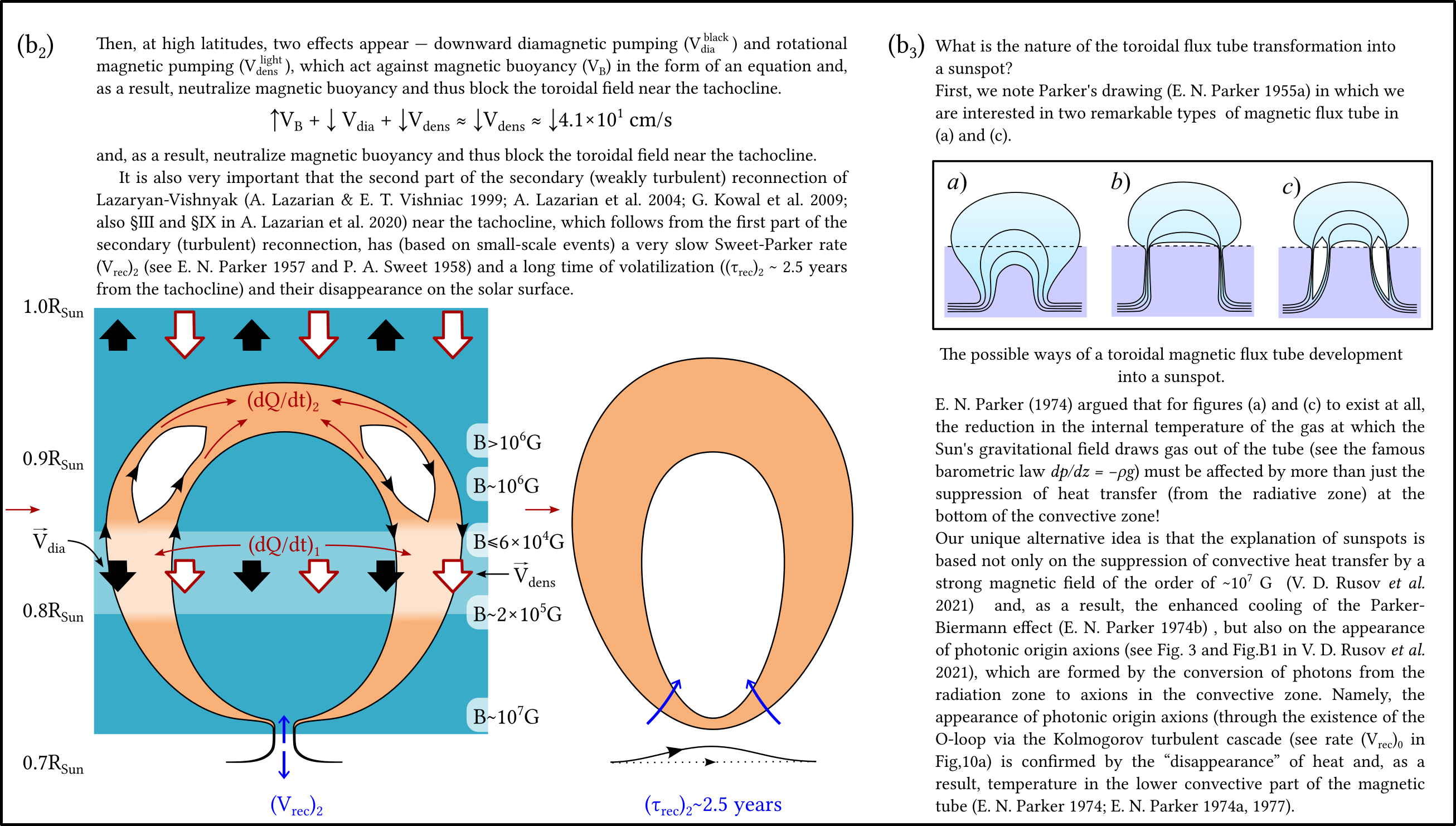}

\vspace{3mm}

\includegraphics[width=\linewidth]{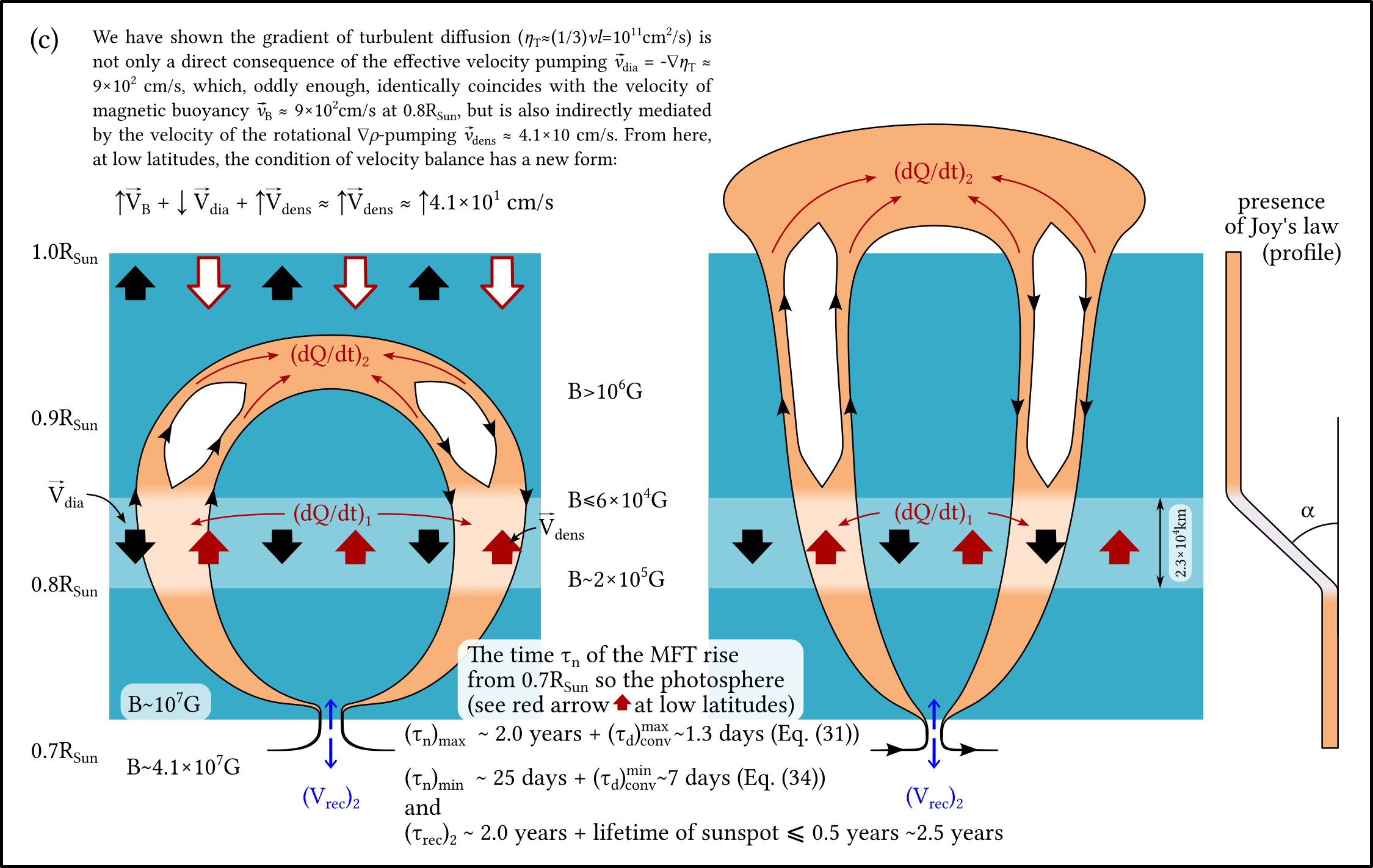}
\end{center}
\caption{(cont.) An alternative theory of magnetic flux tubes in strong fields by 
means of axion origin photons associated with primary and secondary 
reconnections. [Ref: A.~Lazarian \& E.~T.~Vishniac~(1999)~\cite{Lazarian1999}; A.~Lazarian~\textit{et~al.}~(2004)~\cite{Lazarian2004}, (2020)~\cite{Lazarian2020}; G.~Kowal~\textit{et~al.}~(2009)~\cite{Kowal2009};  P.~A.~Sweet~(1958)~\cite{Sweet1958};
E.~N.~Parker~(1955a)~\cite{Parker1955a}, (1957)~\cite{Parker1957}, (1974)~\cite{Parker1974b}, (1974a)~\cite{Parker1974c}, (1974b)~\cite{Parker1974a}, (1977)~\cite{Parker1977}; V.~D.~Rusov~\textit{et~al.}~(2021)~\cite{RusovDarkUniverse2021}]}
\end{figure*}

In this regard, we consider turbulence with quasi-isotropic spectral
tensor~\cite{Kichatinov1987,Kichatinov1991}, which is certainly the simplest
representation for inhomogeneous turbulence (see Eq.~(2.12) 
in~\cite{Kichatinov1992}). As a consequence, the information on the spectral 
properties of turbulence (given by Eqs.~(2.12)-(2.16)
from~\cite{Kichatinov1992}) is sufficient to reduce the expression for the 
average electromotive force $\varepsilon$ (see Eq.~(2.1)~\cite{Kichatinov1992})
to its traditional form, where only integrations over the wave number $k$ and 
frequency $\omega$ remain. After such shortening it is possible to get (see 
Eq.~3.1 in~\cite{Kichatinov1992})
\begin{equation}
\vec{F} = (\vec{v}_{dia} + \vec{v}_{dens}) \times \vec{B}
\label{eq07-69}
\end{equation}

\noindent
with the speed of turbulent diamagnetic transfer

\begin{equation}
\vec{v}_{dia} = -\nabla \int \limits _{0} ^{\infty} \Re _{dia} (k, \omega, B)
\frac{\eta k^2 q (k, \omega, x)}{\omega ^2 + \eta ^2 k^4} dk d\omega ,
\label{eq07-70}
\end{equation}

\noindent
where $q$ stands for the local velocity spectrum, and the rate of rotational magnetic advection caused by the vertical heterogeneity of the fluid density in the convective zone, i.e. the magnetic $\nabla \rho$-pumping effect,

\begin{equation}
\vec{v}_{dens} = \frac{\nabla \rho}{\rho} \int \limits _{0} ^{\infty} \Re _{dens} (k, \omega, B)
\frac{\eta k^2 q (k, \omega, x)}{\omega ^2 + \eta ^2 k^4} dk d\omega .
\label{eq07-71}
\end{equation}

\noindent
The effective speeds $v_{dens}$ and $v_{dia}$ are consequences of the 
non-uniformity of density and of turbulence intensity, respectively, where the 
latter is attributed to the known diamagnetic pumping.
The velocities (\ref{eq07-71}) and (\ref{eq07-70}) depend on the magnetic field
through the kernels $\Re_{dens}$ and $\Re_{dia}$.

We are interested in the problem of reconstructing a strong toroidal field of
flux tubes $\sim 10^7 ~G$ (see Figs.~\ref{fig-meridional-cut}a,b for the 
magnetic tube profile, and Figs.~\ref{fig-prim-sec-reconnection}a,b,c for the 
magnetic tube in full face), transform the regions of the average magnetic 
field $B \sim 2 \cdot 10^5 ~G$ in the convective zone (see 
Figs.~\ref{fig-prim-sec-reconnection}b,c), and thereby allow the organization
of the amazing balance between the magnetic buoyancy, turbulent diamagnetism,
and the rotationally modified $\nabla \rho$-effect.

In this regard, we first consider the physics of linear magnetic diffusivity, 
which is known to govern two processes: macroscopic turbulent diamagnetism and 
rotational $\nabla \rho$-pumping. This is because the standard theory of the 
turbulent mixing length $l$ and the characteristic velocity of the dominant 
eddies $u$ yields a diffusion coefficient $\eta_T = (1/3) u l$, that depends 
directly on the diamagnetic pumping rate $v_{dia} = -\nabla \eta_T/2$ and, 
indirectly, on the rotational $\nabla \rho$-pumping $v_{dens}$.

Here we assumed that, according to 
Figs.~\ref{fig-prim-sec-reconnection}b$_2$,c, the radial profile of turbulent 
diffusion is a smooth-convex function with a maximum, 
$\eta_T \approx 10^{11} ~cm/s$, approximately in the middle of the CZ at a 
depth of $z = 1.30 \cdot 10^5 ~km$ at $\sim 0.8 R_{Sun}$, which is described by
the distance of $2.3 \cdot 10^4 ~km$ between $0.8 R_{Sun}$ and near 
$0.85 R_{Sun}$ (see Figs.~\ref{fig-prim-sec-reconnection}b$_2$,c) and with the
depth of the bottom of the convective zone $z_0 \approx 1.84 \cdot 10^5 ~km$  
(see \cite{Stix2002,Kippenhahn2012,Schumacher2020}).  

In order to consider the balance of the three velocities $v_B$, $v_{dia}$, and
$v_{dens}$, it is necessary to apply the widely used approximation of the 
mixing length (see 
e.g.~\cite{BohmVitense1958,Bradshaw1974,Gough1977a,Gough1977b}),
which, according to~\cite{Kichatinov1991}, fully satisfies this goal. This 
approximation will be understood as the replacement of nonlinear terms along 
with time derivatives in the equations for fluctuating fields by means of 
$\tau$-relaxation terms, i.e. instead of equations (3.1) and (3.9) 
from~\cite{Kichatinov1991}, we now have the equation of the radial speed of the 
toroidal field in the convection zone

\begin{equation}
v_{dens}^{light} = \tau \langle u^2 \rangle ^{\circ}
\frac{\nabla \rho}{\rho} \left[ \phi_2 (\hat{\Omega}) - \cos ^2 \theta \cdot 
\phi_1 (\hat{\Omega}) \right] \approx
\label{eq07-75}
\end{equation}

\begin{equation}
\approx 6 v_p \left[ \phi_2 (\hat{\Omega}) - \cos ^2 \theta \cdot 
\phi_1 (\hat{\Omega}) \right] =
\label{eq07-76}
\end{equation}

\begin{equation}
= - \frac{3 \kappa g}{(\gamma - 1)c_p T}
\left[ \phi_2 (\hat{\Omega}) - \cos ^2 \theta \cdot \phi_1 (\hat{\Omega}) \right] ,
\label{eq07-77}
\end{equation}

\noindent
and \underline{\textbf{firstly}}, it can be shown that at 
$\eta _T \approx 10^{11} ~cm/s$, we have
\begin{equation}
v_{dens}^{light} \sim 3 \times \eta_T \times 
g_{0.8 R_{Sun}} / (\gamma - 1) c_p T_{0.8 R_{Sun}} \approx 4.1 \cdot 10^1 ~cm/s ,
\label{eq08-2.29}
\end{equation}

\noindent
where $\tau \approx l / (\langle u^2 \rangle ^{\circ})^{1/2}$ is a typical 
lifetime of a convective eddy; $l$ is the mixing length; 
$\langle u^2 \rangle ^{\circ}$ is the mean intensity of fluctuating velocities 
for original turbulence; $\theta$ is the latitude (scalar); 
$e_r \nabla \rho / \rho = -e_r g / [(\gamma - 1) c_p T]$ , where $e_r$ is the 
radial unit vector; $T \cong 1.352 \cdot 10^6 ~K$ is the temperature at 
$0.8 R_{Sun}$~\cite{Bahcall1992}; $g = g_0 (R_{Sun}/r)$ is the gravity, where
$g_0 = 2.74 \cdot 10^4 ~cm/s^2$ is the surface value of solar gravity, and 
$g_{0.8 R_{Sun}} = 4.22 \cdot 10^4 cm/s^2$ at $0.8 R_{Sun}$; 
$c_p = 3.4 \cdot 10^8 ~cm^2 s^{-2} K^{-1}$ (fully ionized hydrogen) is the 
specific heat at constant pressure; $\gamma = 5/3$ is the ratio of specific 
heats $c_p / c_V$; $3 \kappa = \tau \langle u^2 \rangle ^{\circ}$, where 
$\kappa \equiv \eta _T \approx 10^{11} ~cm^2/s$ is turbulent diffusivity 
coefficient, and the mixing length relation 
$\langle u^2 \rangle ^{\circ} = - \nabla \Delta T l^2 g / (4 T)$, where the 
$\nabla \Delta T$ is superadiabatic temperature gradient; 
$v_p = (1/6) \tau \langle u^2 \rangle ^{\circ} (\nabla \rho / \rho)$ is the 
velocity of the magnetic field transfer caused by the density gradient (see 
Eq.~(36) in~\cite{VainshteinKichatinov1983}); 
$\hat{\Omega} = Co = 2 \tau \Omega$ is the Coriolis number (reciprocal of the 
Rossby number), where $\Omega$ is the rotation speed, $\tau$ is the turnover 
time, and the functions

\begin{equation}
\phi _n (\hat{\Omega}) = (1/8) I_n (\Omega, k, \omega)
\label{eq07-78}
\end{equation}

\noindent
(see $I_1$ and $I_2$ in Eqs.~(3.12) and~(3.21) in \cite{Kichatinov1991}) are

\begin{equation}
\phi _1 (\hat{\Omega}) = \frac{1}{4 \hat{\Omega}^2} \left[
-3 + \frac{\hat{\Omega}^2 + 3}{\hat{\Omega}} \arctan \hat{\Omega} \right] ,
\label{eq07-79}
\end{equation}

\begin{equation}
\phi _2 (\hat{\Omega}) = \frac{1}{8 \hat{\Omega}^2} \left[
1 + \frac{\hat{\Omega}^2 - 1}{\hat{\Omega}} \arctan \hat{\Omega} \right]
\label{eq07-80}
\end{equation}

\noindent
(see also analogous Eq.~(19) and Fig.~2 in~\cite{Kitchatinov2016}), which 
describe the rotational effect on turbulent convection.

From here, we obtain an the estimate of the Coriolis number for solar 
convection in the deep layer, which should be $\hat{\Omega} \approx 20$. 
By virtue of the equations (\ref{eq07-78})-(\ref{eq07-80}) and the function
$\phi_n$ (see Eq.~(\ref{eq07-78})) for the value $\hat{\Omega} \approx 20$ (see
from $0.8 R_{Sun}$ in Fig.~\ref{fig-prim-sec-reconnection}b$_2$,c), the 
following values are adopted:

\begin{equation}
\phi_1 \cong 0.0171 , ~~~ \phi_2 \cong 0.0098 ,
\label{eq07-81}
\end{equation}

\noindent
at which the radial velocity $v_{dens}^{light}$ in Eq.~(\ref{eq07-75}) of 
toroidal field transport changes the sign at the latitude 
$\theta ^{*} = \arccos \sqrt{\varphi_2 / \varphi_1} \cong 41^{\circ}$, being 
negative (downward) for $\theta > \theta ^{*}$ and positive (upward) for 
$\theta < \theta ^{*}$. Using Eq.~(\ref{eq07-77}), we find that the value of 
the radial velocity of toroidal field transport in the convective zone 
$v_{dens}^{light}$ (see Eqs.~(\ref{eq07-75})-(\ref{eq07-77})) near low 
latitudes (e.g. $\theta ^{*} = \arccos (0.985) \cong 10^{\circ}$; see also 
Fig.~\ref{fig-meridional-cut}b) is almost identical to the value of 
the velocity (\ref{eq08-2.29}), which was previously calculated using the 
toroidal field $B \approx 2 \cdot 10^5 ~G$ at $0.8 R_{Sun}$ (see 
Eq.~(\ref{eq08-2.6a}) and Fig.~\ref{fig-prim-sec-reconnection}b,c) in the 
$\Re_{dens} (\beta, \varphi)$ (see Eq.~(3.13) in \citep{Kichatinov1992}).

\underline{\textbf{Secondly}}, we get diamagnetic pumping rate
\begin{equation}
v_{dia} = -\nabla \eta _T / 2 \approx 1.7 \cdot 10^2 ~cm/s .
\label{eq08-2.33}
\end{equation}

\underline{\textbf{Thirdly}}, in addition to the radial profile of turbulent 
diffusion $\eta _T \approx 10^{11} ~cm^2/s$, which is related to the 
diamagnetic pumping velocity $v_{dia} \approx 1.7 \cdot 10^2 ~cm/s$ and the 
distance $2.3 \cdot 10^4 ~km$ between $0.8 R_{Sun}$ and near $0.85 R_{Sun}$ 
(see the ``green'' line of the turbulent-toroidal magnetic field in 
Fig.~\ref{fig-meridional-cut}b and Fig.~\ref{fig-prim-sec-reconnection}b,c), we
need to calculate the magnetic buoyancy velocity $v_B$, which is associated 
with a super-uniformly distributed field strength $B \sim 2\cdot 10^5 ~G$ (see 
Fig.~\ref{fig-meridional-cut}b and Fig.~\ref{fig-prim-sec-reconnection}b,c).
It should be remembered that for buoyant flux tubes with super-equipartition 
field strength 
$B \sim 2 \cdot 10^5 ~G \geqslant (H_p/a)^{1/2} B_{eq} \approx 3.16 B_{eq}$ 
with $a \approx 0.1 H_p$, where, as Yuhong Fan wrote (see Section~7 
in~\cite{Fan2021}), ``...$B_{eq} \approx 6.1 \cdot 10^4 ~G$ is in equipartition
with the kinetic energy density of the convective downflows, the magnetic 
buoyancy of the tubes dominates the hydrodynamic force from the convective 
downflows and the flux tubes can rise unaffected by convection'', but 
significantly affects the turbulent diamagnetic pumping!

For this, using our so-called universal van~Ballegooijen-Fan-Fisher model with 
equations (\ref{eq07-44})-(\ref{eq08-2.8}) with the values of the magnetic flux
tube with $\sim 2\cdot 10^5 ~G$, the pressure of 
$1.665 \cdot 10^{13} ~erg/cm^3$ at $0.8 R_{Sun}$ and the rate of radiative 
heating $(dQ/dt)_1 \approx 29.7 ~erg \cdot cm^{-3} \cdot s^{-1}$, we obtained 
the magnetic buoyancy velocity $v_B \approx 1.7 \cdot 10^2 ~cm/s$, which, 
surprisingly, exactly coincides with the diamagnetic pumping velocity 
$v_{dia} \approx 1.7 \cdot 10^2 ~cm/s$.

\begin{equation}
v_B \approx 1.7 \cdot 10^2 ~cm/s ,
\label{eq08-2.34}
\end{equation}

\noindent
which, surprisingly, exactly coincides with the diamagnetic pumping rate 
$v_{dia} \approx 1.7 \cdot 10^2 ~cm/s$.

From here, a certain balance of three velocities, $\vec{v}_B$, $\vec{v}_{dia}$ 
and $\vec{v}_{dens}$, arises, which at 
$v_B = v_{dia} \approx 1.7 \cdot 10^2 ~cm/s$ provide both the 
process of blocking (via $\vec{v}_{dens}$ at $0.8 R_{Sun}$) magnetic 
buoyancy at high latitudes
\begin{equation}
\uparrow v_B + \downarrow v_{dia}^{black} + \downarrow v_{dens}^{light} =
\downarrow v_{dens}^{light} \cong  \downarrow 4.1 \cdot 10^1 ~cm/s ,
\label{eq08-2.35}
\end{equation}

\noindent
and the process of raising magnetic buoyancy (via $\vec{v}_{dens}$ at 
$0.8 R_{Sun}$) to the surface of the solar photosphere at low latitudes
\begin{equation}
\uparrow v_B + \downarrow v_{dia}^{black} + \uparrow v_{dens}^{light} =
\uparrow v_{dens}^{light} \cong  \uparrow 4.1 \cdot 10^1 ~cm/s .
\label{eq08-2.36}
\end{equation}

One of the main tasks was related to the solution of the maximum 
near-equatorial time $(\tau _n)_{max}$ of the rise of the magnetic flux tube in
the region between $0.8 R_{Sun}$ and near $0.85 R_{Sun}$ (see 
Fig.~\ref{fig-meridional-cut}b and Fig.~\ref{fig-prim-sec-reconnection}b,c), 
which is associated with the distance of turbulent diffusion 
$l_{diffus} = 2.3 \cdot 10^4 ~km$ and the speed of rotational 
$\nabla \rho$-pumping $v_{dens} \approx 4.1 \cdot 10 ~cm/s$:

\begin{align}
(\tau _n)_{max} & = \frac{l_{diffus}}{v_{dens}} \approx 0.56 \cdot 10^8 ~s \sim 2~years , \nonumber \\
(\tau _n)_{min} & = \frac{l_{diffus}}{v_{dens}} \approx 0.56 \cdot 10^6 ~s \leqslant 25~days .
\label{eq08-2.37}
\end{align}

When the toroidal fields of magnetic tubes rising from the tachocline to the 
solar surface, but without the participation of the turbulent diffusion 
distance (see Eq.~\ref{eq07-68}), have strong fields of the order of 
$> 2 \cdot 10^5 ~G$ (see Fig.~\ref{fig-meridional-cut}a and 
Fig.~\ref{fig-prim-sec-reconnection}a), then they
\begin{equation}
v_B (> 2\cdot 10^5 ~G) \gg v_{dia} + v_{dens} ,
\label{eq08-2.38}
\end{equation}

\noindent
suppress the existence of the velocities $\vec{v}_{dia}$ and $\vec{v}_{dens}$! 
This means that the near-equatorial time $(\tau_n)_{max}$ of the magnetic flux 
tube rise, equal to 2 years, neglects the time of convective heating of the 
flux tube, which corresponds to only a few days (see analogue of 
$(\tau_d)_{conv}^{max}$ in Eq.~\ref{eq08-2.19a2} and $(\tau_d)_{conv}^{min}$ 
in Eq.~\ref{eq08-2.20c}).

The most interesting thing is that, in addition to the sunspots after the rise
of the magnetic tube with $B \sim 10^7 ~G$ at all latitudes (see 
Fig.~\ref{fig-meridional-cut}a and Fig.~\ref{fig-prim-sec-reconnection}a) and 
the sunspots after the rise of the magnetic tube with 
$B \leqslant 2\cdot 10^5 ~G$ only at low latitudes (see 
Fig.~\ref{fig-meridional-cut}b and Fig.~\ref{fig-prim-sec-reconnection}c), 
which are the source of the double maxima of the 11-year solar cycles (see the
full experimental evidence in Section~\ref{sec-double-maxima}), we are 
interested in the problem or, more precisely, the physics of the tilt angle of 
Joy's law~\cite{Ivanov2012,DasiEspuig2010} and how it is related to the 
magnetic $\nabla \rho$-pumping?

From here, this raises an unexpected question: In what way, in addition to the 
visible magnetic tube at strong fields at all latitudes (see 
Fig.~\ref{fig-meridional-cut}a and Fig.~\ref{fig-prim-sec-reconnection}a), the 
``invisible'' magnetic tube at high latitudes (see 
Fig.~\ref{fig-prim-sec-reconnection}b$_2$) and the ``visible'' magnetic tube at
low latitudes (see Fig.~\ref{fig-prim-sec-reconnection}c), depending on the 
time $(\tau _{rec})_2 \sim 2.5 ~years$ of reconnection of the horizontal field 
in the tachocline (which is equal to the sum of 
$(\tau_n)_{max} \approx 2 ~years$ and the lifetime of sunspots 
($\leqslant 0.5 ~years$) from several days to several months 
(see~\cite{DSilvaChoudhuri1993,Fan1993,Fan1994}) surprisingly, practically does not
depend on the direction in the radial plane down or up?

The answer, surprisingly, is not very simple. We know that the vertical 
compression (see Fig.~\ref{fig-meridional-cut}a and 
Fig.~\ref{fig-prim-sec-reconnection}b$_1$) of the descending nearly empty 
anchored magnetic tube in strong fields of $\sim 10^7 ~G$ (and via convective 
heating $(dQ/dt)_2$ is associated with a horizontal ``stretching'' or expansion
of this magnetic tube near the tachocline, where the counter-movement of 
magnetic fields leads to the emergence of a collisionless interaction with fast
(opposite) velocities $(V_{rec})_1$ (and times $(\tau _{rec})_1$ ($<$10 days)) 
of the first part of the fast secondary (turbulent) reconnection (see 
Fig.~\ref{fig-prim-sec-reconnection}b$_1$). When the vertical compression leads
to horizontal stronger ``stretching'' of the magnetic tube in weak fields of 
$\geqslant 2\cdot 10^5 ~G$ (and through radiative heating $(dQ/dt)_1$, then the
counter motion of magnetic fields (see $(V_{rec})_1$ for 
Fig.~\ref{fig-prim-sec-reconnection}b$_1$) leads to the emergence of the second
part of the secondary (turbulent) reconnection with a very slow Sweet-Parker 
speed $(V_{rec})_2$ (see Fig.~\ref{fig-prim-sec-reconnection}b$_2$; also~\cite{Lazarian1999,Lazarian2004,Kowal2009,Lazarian2020,Parker1957,Sweet1958}). 
This means that the emerging weak fields of $\geqslant 2 \cdot 10^5 ~G$ at 
distances from $0.80 R_{Sun}$ to near $0.85 R_{Sun}$ are a remarkable source of
rotating magnetic $\nabla \rho$-pumping with the rise time of the magnetic tube
$(\tau _n)_{max} \sim 2.0 ~years$ from the tachocline to the solar surface (see
Fig.~\ref{fig-prim-sec-reconnection}c), and the lifetime of sunspots 
($\leqslant 0.5 ~years$), and consequently, the time of the magnetic tube 
evaporation $(\tau _{rec})_2 \leqslant 2.5 ~years$ and their disappearance on 
the solar surface.

Since the study of the tendency of the tilt angle of Joy's law is very 
important for understanding the evolution of the solar magnetic field, then 
unlike the heavy calculations of theoretical estimates of the averaged angle 
tilt of sunspots with increasing latitude, we can summarize our results (see 
Eqs.~(5)-(9) in~\cite{DSilvaChoudhuri1993}), which can simultaneously be 
expressed in the form of simple and understandable physics of Joy's law. It is
known that, according to~\cite{DSilvaChoudhuri1993} and~\cite{Fan1993}, the 
Coriolis force is proportional to the magnetic buoyancy velocity $v_B$, which 
can be estimated taking into account the balance between the buoyancy force 
(the term on the left) and the drag force (left term) and the drag force (right
term): 
\begin{equation}
\frac{B^2}{8\pi H_p} = C_D \frac{\rho_{ext} v^2_B}{\pi a} ,
\label{eq08-2.54}
\end{equation}

\noindent
where the buoyancy velocity $v_B$ is related to the radius $a$ of the 
cross-section of the magnetic tube.

Taking into account Eqs.~(18) and (19) by~\cite{Fan1994} and using the 
theoretical values of the tilt angle $\alpha$, the latitude angle $\theta$ and 
the radial parameter $\xi = r / R_{Sun} \sim 0.80$, one can easily obtain a 
simplified equation in the following form:
\begin{equation}
\sin {(tilt)} \propto \frac{\sqrt{a(\xi, \theta)}}{B} \sin {(latitude)} .
\label{eq08-2.55}
\end{equation}

This means that looking at equation (\ref{eq08-2.55}), we understand that when
magnetic tubes contain relatively large radii $a$ (due to radiative heat 
transfer $(dQ/dt)_1$ (see Fig.~\ref{fig-prim-sec-reconnection}b$_2$,c) and low
magnetic fields ($\leqslant 10^5 ~G$), then the tilt angle of Joy's law are 
born from here, and when magnetic tubes have practically zero radii $a$ (due to
convective heat transfer $(dQ/dt)_2$ (see 
Fig.~\ref{fig-prim-sec-reconnection}b$_2$,c) and strong high magnetic fields 
($> 10^5 ~G$), then the tilt angle of Joy's law disappears!

It can be said in a completely different way. If radiative heat transfer 
$(dQ/dt)_1$ (see Fig.~\ref{fig-prim-sec-reconnection}b$_2$,c) or the presence 
of magnetic $\nabla \rho$-pumping is used in magnetic tubes, then, as usual, 
the tilt angles of Joy's law appear in sunspots, when convective heat transfer 
$(dQ/dt)_2$ or the disappearance of magnetic $\nabla \rho$-pumping (see 
Fig.~\ref{fig-prim-sec-reconnection}b$_2$,c) is used in magnetic tubes, then, 
strangely enough, the tilt angles of Joy's law disappear in sunspots!

In this regard, we are interested in an intriguing question: Where and how do 
sunspots appear, in which, according to~\cite{Gnevyshev1963}, the maxima of the
double peak in the sunspot cycle appear? And how does the first maximum -- the 
maximum number of spots -- physically differ from the second -- the maximum of 
large spots?

Below we will show that, firstly, the first maximum differs from the second 
maximum by being shifted in time by 2 years (see text after 
Eq.~(\ref{eq08-2.38})) later than the main (first!) maximum of the solar 
activity cycle. Secondly, according to equation (\ref{eq08-2.54}), when the 
flux tube has a large radius and a low magnetic field, then this magnetic tube,
surprisingly, is the source of the formation of the tilt angle (via the 
Coriolis force, and not necessarily on the surface of the photosphere (see 
Fig.~\ref{fig-prim-sec-reconnection}c)) and a large area of the spot on the 
surface of the Sun. Conversely, when the flux tube has a small radius and a 
high magnetic field, then the magnetic tube, on the one hand, is not connected
to the source of the formation of the tilt angle of Joy's law (through the 
Coriolis force), and on the other hand, surprisingly, has a small spot area on
the surface of the Sun.

This raises a second intriguing point. One of the first solutions in our 
section is secondary reconnection of the MFT, which involves a turbulent 
diffusivity of $10^{11} ~cm/s$ and a tilt angle of Joy's law between magnetic
fields ranging from $2 \cdot 10^5 ~G$ (at $0.8 R_{Sun}$) to $6\cdot 10^4 ~G$ 
(at $0.85 R_{Sun}$). Moreover, due to the relationship between turbulent diffusivity and the tilt angle of Joy's law, the direction of magnetic $\nabla \rho$-pumping (upward or downward (see Fig.~\ref{fig-prim-sec-reconnection}b$_2$,c)) is sensitive to the sign of the multiplier in equations (\ref{eq07-79})-(\ref{eq07-79}), which depends on the polar angle (co-latitude) and the behavior of the Coriolis number functions in the convective zone.

It is crucial that the existence of secondary MFT reconnection generates a 
secondary maximum in the 11-year sunspot cycle. This is because when the 
primary reconnection, in which the buoyancy of a practically empty magnetic 
flux tube at $\sim 10^7 ~G$ rises from the tachocline to the photosphere for
$\sim 2$ to $10$~days, transitions to secondary reconnection, the latter is 
associated with a buoyancy of practically MFT from $\sim 6 \cdot 10^4 ~G$ from 
$0.85 R_{Sun}$ (see Fig.~\ref{fig-prim-sec-reconnection}c)) to the photosphere.
From this we understand that since the lifetime of $\sim 2$~years for the 
rotating magnetic $\nabla \rho$-pumping from $2 \cdot 10^5 ~G$ at $0.8 R_{Sun}$
(see Fig.~\ref{fig-prim-sec-reconnection}c) to $6\cdot 10^4 ~G$ is related to 
the lifetime of the order of $\sim 10$~days for the buoyancy of almost MFT from
$\sim 6\cdot 10^4 ~G$ to the photosphere, this means that the secondary 
reconnection of MFT from the tachocline to the photosphere is also equal to 
$\sim 2$~years.

In other words, the connection between primary and secondary reconnections (or 
first and second maxima) is the connection of MFTs between lifetimes of the 
order of $\sim 2$ to $10$~days and lifetimes of the order of $\sim 2$~years, 
which is preserved by 11-year sunspot cycles without any dynamo.

\subsubsection{Double maxima of 11-year sunspot cycles and the Gnevyshev gap}
\label{sec-double-maxima}

\begin{figure}
\begin{center}
\includegraphics[width=\linewidth]{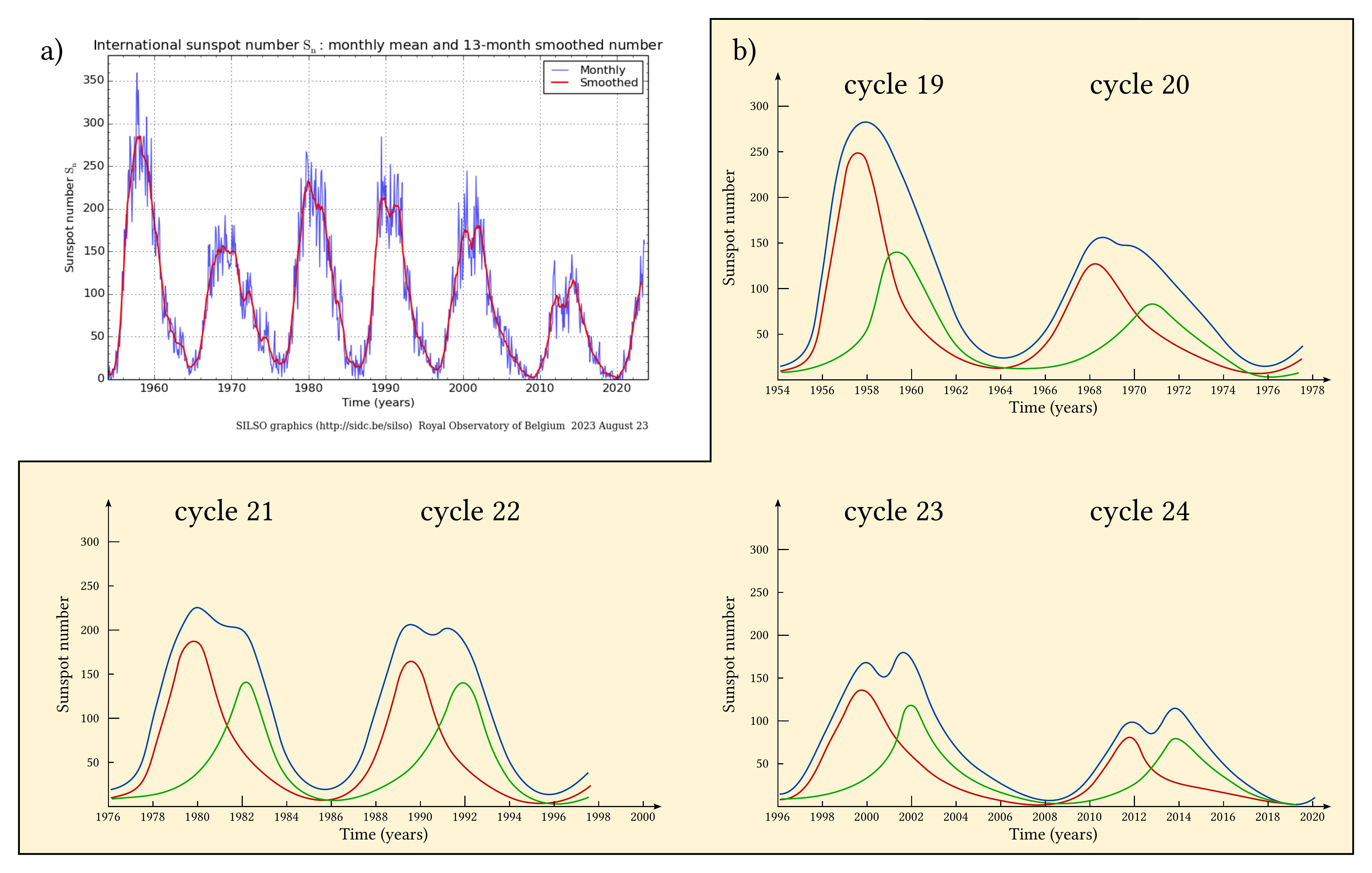}
\end{center}
\caption{Variation in the number of sunspots for six cycles from 19 to 24 
associated with double maxima of 11-year solar cycles: 
(a) solid red lines of sunspot cycles -- smoothed values 
(\url{https://sidc.be/SILSO}) Royal Observatory of Belgium 2023, August 23.
(b) theoretical assessment of double maxima of 11-year solar cycles: solid line
 -- ``smoothed'' blue values of sunspot cycles, in which the sum of the curves
with the first maximum in the form of a ``red'' value and the second maximum in
the form of a ``green'' value almost exactly coincides with a solid ``blue'' 
line of the corresponding cycle.}
\label{fig-solar-cycles-sunspots}
\end{figure}

The phenomenon of a double peak in the sunspot cycle was discovered by Mstislav
Gnevyshev in 1963~\cite{Gnevyshev1963} during a study of coronal emission and 
areas of active formations in the 11-year cycle 19~\cite{Gnevyshev1963}. First,
Gnevyshev analyzed the evolution of the annual average intensity of the green 
coronal spectral line $\lambda 530.3$ (which serves as a direct indicator of 
magnetic activity and coronal heating) in five-degree heliolatitude intervals 
during the cycle 19 and discovered two maxima in the intensity of the corona. 
According to data in~\cite{Antalova1965,Gnevyshev1967,Gnevyshev1977}, the first
maximum of the area of sunspot groups coincides with the main maximum of the 
11-year cycle for Wolf numbers, while the second maximum is associated with an
increase in the number of large sunspots (the so-called maximum of sunspot
intensities)~\cite{Vitinskij1986,Kopecky1969}. In other words, the first 
maximum is the maximum in the number of sunspots, and the second is the maximum
of large sunspots~\cite{Kopecky1969}. The more large spots in the cycle, the 
more distinct the two peaks. These features were later confirmed and for the 
cycle 20 (see Fig.~\ref{fig-solar-cycles-sunspots}), including for the solar 
northern and southern hemispheres separately~\cite{Gnevyshev1963}.

In contrast to the ``new solar dynamo era'' (see e.g.
\cite{Krivodubskij2005,Fan2021,Zhang2022,Zhang2023,Charbonneau2023,Vasil2024,Kapyla2023,Chatterjee2026}),
we used our alternative theory of magnetic flux tubes at strong fields via 
photons of axion origin. The essence of double maximum 11-year sunspot cycles, 
the source of which is the generation of a magnetic field near the bottom of 
the convective zone and the subsequent rise of the field from the deep layers 
to the surface in the photosphere, manifests itself at strong fields only (see 
Fig.~\ref{fig-prim-sec-reconnection}a, 
Fig.~\ref{fig-prim-sec-reconnection}b$_2$,c).
This means that the two maxima are related as follows: the first maximum (the
primary reconnection of the empty MFT with $10^7 ~G$ at $0.72 R_{Sun}$; see 
Figs.~\ref{fig-meridional-cut}a, \ref{fig-prim-sec-reconnection}a),
which generates a very ``fast'' velocity to the solar surface (on the order 
of several days; see Eq.~(\ref{eq08-2.19a2}) $\sim 1-2 ~days$) is transformed 
into a secondary maximum (the secondary reconnection of the empty MFT with 
$10^6 ~G$ at $0.85 R_{Sun}$; see Figs.~\ref{fig-meridional-cut}a, 
\ref{fig-prim-sec-reconnection}a), which then generates a ``very slow'' 
velocity to the solar surface (on the order of two years; see 
Eq.~(\ref{eq08-2.37}) of
$(\tau_n)_{max} \sim 2.0 ~years$ for 
Fig.~\ref{fig-prim-sec-reconnection}b$_2$,c). 

In other words, the ``very slow'' second maximum depends only on two 
interrelated causes: the very slow secondary reconnection (see 
Fig.~\ref{fig-prim-sec-reconnection}b$_2$,c) of the magnetic flux tube, which 
depends both at a strong magnetic field of $>2 \cdot 10^5 ~G$, having two
sections between $0.7 R_{Sun}$ and near $0.80 R_{Sun}$, and between 
$0.85 R_{Sun}$ and the photosphere (see 
Fig.~\ref{fig-prim-sec-reconnection}b$_2$,c), and, surprisingly, at a weak
magnetic field of $<2 \cdot 10^5 ~G$, where the so-called effects of
turbulent diamagnetic transport and rotating $\nabla \rho$-pumping appear for 
the first time, which provide both the process of blocking magnetic buoyancy 
at high latitudes (see Fig.~\ref{fig-prim-sec-reconnection}b$_2$), and at 
near-equatorial latitudes predetermines the process of dominance of the 
Coriolis force with rotating $\nabla \rho$-pumping with 
$<2 \cdot 10^5 ~G$ at a distance $2.3 \cdot 10^4 ~km$ between 
$0.80 R_{Sun}$ and near $0.85 R_{Sun}$ (see 
Fig.~\ref{fig-prim-sec-reconnection}c), which ultimately generates a ``shift''
of the arcuate ascending loop in the form of the tilt angle of Joy's law on the
solar surface (see Fig.~\ref{fig-prim-sec-reconnection}c).

In this case, the key role in the proposed mechanism of double maxima is played
by two ``waves'' of toroidal fields, in which the first wave, with a strong 
field of $\sim 10^7 ~G$ and without the participation of secondary 
reconnection, first arises at all latitudes, and then, as a consequence, 
instead of the first wave, a second wave appears with a weak magnetic field of
$<2 \cdot 10^5 ~G$ between $0.80 R_{Sun}$ and $0.85 R_{Sun}$, which is obtained
from the same primary wave, but with the participation of secondary 
reconnection, only at the latitudes of the equatorial surface of the Sun.

That is to say, the first maximum is the maximum of the number of spots that 
are sources of the magnetic flux tube (anchored in the overshoot tachocline) 
with $B \sim 10^7 ~G$ and with the participation of primary reconnection (see
Fig.~\ref{fig-Kolmogorov-cascade} and Fig.~\ref{fig-lower-reconnection}a; also 
Fig.~\ref{fig-meridional-cut}a), and the second maximum is the maximum of large
spots that are the result of the appearance of magnetic flux tubes with 
$B \sim 10^7 ~G$ and with the participation of secondary reconnection (see 
Fig.~\ref{fig-lower-reconnection}b and 
Fig.~\ref{fig-prim-sec-reconnection}b$_2$,c). From here, we are interested, in 
fact, in the very important secondary magnetic reconnection (see 
Fig.~\ref{fig-meridional-cut}b and Fig.~\ref{fig-prim-sec-reconnection}b$_2$,c)
during the rise of the magnetic flux tube in the solar photosphere, which is 
one of the two sources of the observed double maxima of one sunspot cycle -- it
shifts by $\sim 2.5~years$ between the primary (see the ``red'' curves in 
Fig.~\ref{fig-solar-cycles-sunspots}b) and secondary (see the ``green'' curves
in Fig.~\ref{fig-solar-cycles-sunspots}b) sunspot maxima (see ``smoothed'' blue
curves in Fig.~\ref{fig-solar-cycles-sunspots}b). Moreover, it is the secondary
maximum of sunspots that is also associated with both observational data of the
tilt angle of Joy's law (see Fig.~\ref{fig-prim-sec-reconnection}c from the 
right) and with the disappearance of sunspots on the surface of the Sun (see 
Fig.~\ref{fig-lower-reconnection}c,d).

And finally, we realize that the observed double maxima, for example in cycles
19-24 (see Fig.~\ref{fig-solar-cycles-sunspots}a), also have different ratios 
between the numbers of very ``fast'' primary maxima $(S_n)_{max}^{primary}$ 
(see the red curves in Fig.~\ref{fig-solar-cycles-sunspots}b; also 
Fig.~\ref{fig-meridional-cut}a and Fig.~\ref{fig-prim-sec-reconnection}a) and 
very ``slow'' secondary maxima (see the ``green'' curves in 
Fig.~\ref{fig-solar-cycles-sunspots}b)
\begin{equation}
(S_n)_{max}^{primary} / (S_n)_{max}^{secondary} \approx
\begin{cases}
  2 & at \sim 300 ~sunspots \\
1.5 & at \sim 150 \div 200 \\
  1 & at \sim 100 ~sunspots
\end{cases}
\end{equation}

\noindent
and at the same time have almost the same difference between the secondary 
reconnection time $(\tau _{rec})_2 \leqslant 2.5 ~years$ of the horizontal 
field in the tachocline (which is equal to the sum of 
$(\tau_n)_{max} \approx 2$ and the lifetime of sunspots 
($\leqslant 0.5 ~years$) from several days to several months 
(see~\cite{Solanki2003,Petrovay1997}), which shift by $\sim 2.0 ~years$,
\begin{equation}
(\tau_n)_{max} = (\tau _{rec})_{max}^{secondary} - 
(\tau _{rec})_{max}^{primary} \sim 2.0 ~years
\end{equation}

\noindent
between the ``experimental'' data of double sunspot maxima -- very ``slow''
second maxima $(\tau _{rec})_{max}^{secondary}$ and very ``fast'' primary 
maxima $(\tau _{rec})_{max}^{primary}$ (see 
Fig.~\ref{fig-solar-cycles-sunspots}b).

This raises a valid question: How is the unique source of the butterfly diagram
found only in the form of 11-year variations in sunspot numbers or only in the 
form of alternating 22-year fundamental cycles of solar activity?

Here we show that, as expected, the source of the butterfly diagram is found
only in the 11-year variations of the sunspot numbers 
(see~\cite{Gnevyshev1967}).
Below we present an understanding of the so-called latitude-time interval (or 
gap) between nearby butterflies by the remarkable solar 
physicist Gnevyshev~\cite{Gnevyshev1967}.

To understand the physics of the Gnevyshev gap, we need to briefly describe the
solution of the following interrelated problems:
\begin{itemize}
\item First, these are 11-year variations of sunspots 
(see~\cite{RusovDarkUniverse2021}). A unique result of our model is the fact 
that the 11-year periods, velocities and modulations of the S102 star (see 
Fig.~6b in~\cite{RusovDarkUniverse2021}) near the black hole are a significant 
indicator of the density modulation of the ADM halo in the fundamental plane of
the Galactic center, which closely correlates with the density modulation of 
baryonic matter near the SMBH. If the ADM halo modulations in the GC black hole
lead to ADM density modulations on the solar surface (via vertical density 
waves between the disk and the black hole to the solar neighborhood), then 
there is an ``experimental'' anticorrelation identity between such indicators 
as the 11-year ADM density modulation in the solar interior and the modulations
of solar axions (or photons of axion origin), and equivalently between the 
modulations of solar axions (or photons of axion origin) and the sunspot 
cycles!


\item Secondly, we know why we never see sunspots at the solar poles. The 
answer is quite simple. We know that both the tilt angle between the geometric
and magnetic meridians, which are tilted by 7.25 degrees relative to the 
ecliptic, and the magnetic field of the northern (or southern) part of the 
meridional circulation, which radially corresponds to a length of about 
$\sim 22^\circ$, add up to about $\sim 30^\circ$. It follows clearly from this
that the magnetic field of the meridional circulation part will always slow 
down the magnetic field of the flux tube, which will almost never rise to the
surface of the photosphere. This means that the magnetic fields of the flux 
tubes will only appear below about $90^\circ - 30^\circ \approx 60^\circ$ at
high latitudes. And that is precisely why no one observes sunspots at the poles
of the Sun above $60^\circ$!
 
\item And Thirdly, we can finally observe (see Fig.~6a 
of~\cite{RusovDarkUniverse2021} for Japanese x-ray telescope Yohkoh 
(1991–2001)) and obtain the physics of photons of axion origin, which are 
associated with primary reconnection, only in those sunspots which are 
practically independent of the differential rotation of the Sun (when the
equator rotates faster than the poles) in the butterfly diagram. 
Clearly, in other sunspots which depend exactly on secondary
reconnection of magnetic flux tubes and, as a consequence, on the tilt angle of
Joy's law on the surface of the Sun (see 
Fig.~\ref{fig-prim-sec-reconnection}c), there can never be occurrences of axion
origin photons in sunspots due to the differential rotation of the Sun.
\end{itemize}

\begin{figure}
\begin{center}
\includegraphics[width=\linewidth]{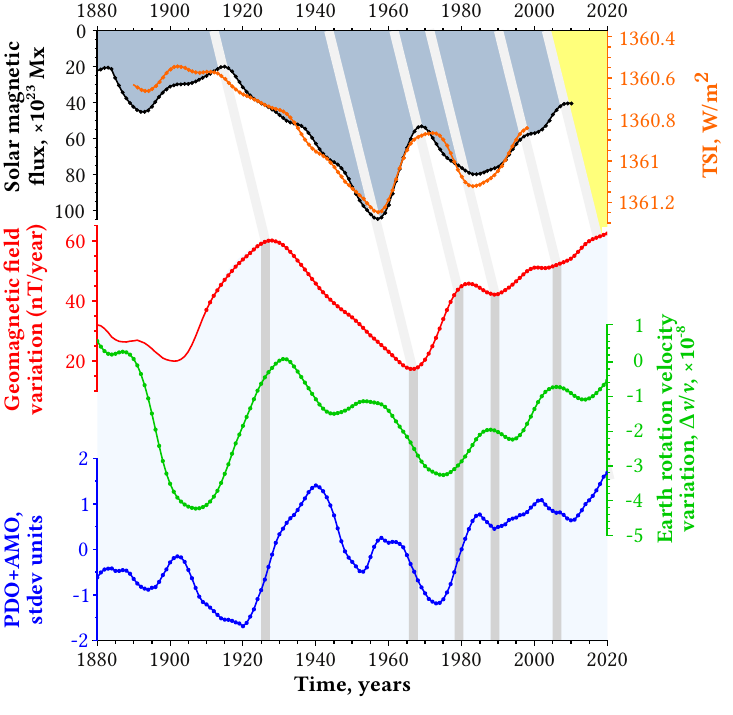}
\end{center}
\caption{The time variations 
of total solar irradiance (TSI; orange curve; see Fig.~9 in~\cite{Pelt2012}),
simulated solar
magnetic flux (black curve; see Fig.~7 in~\cite{Dikpati2008ApJ675}; 
and also Fig.~16 in~\cite{Rusov2013}) (both curves were smoothed with 29-year 
weighted moving average);
the geomagnetic field secular variations (the Y-component; red curve) 
measured at the Eskdalemuir observatory~\cite{Geomagnetism2025};	
variations in the Earth's rotation speed (green curve; joined data from
Fig.~3 in~\cite{Sidorenkov2016}, Fig.~3.9 in~\cite{Sidorenkov2009}, and the 
EOP C01 IAU1980 (1846-now) dataset~\cite{IERS_EOD2025});
the variations of the average global ocean level (Pacific Decadal Oscillation 
(PDO)~\cite{PDO2025} + Atlantic Multidecadal Oscillation (AMO)~\cite{AMO2023}; blue curve) with a trend in mm/years near 60-year oscillation 
(see Fig.~1 in~\cite{Chambers2012}).
The latter three curves were first smoothed using 5-year moving average, 
and then the 11-year moving average. 
All considered time variations between the Sun and the Earth always have a time
interval of approximately $\sim 10\div11$ years. The only serious disagreements
between the reconstructions and observations occur during the Second World War,
especially in the period 1944-1945 (see~\cite{Folland2018}).}
\label{fig-time-variations}
\end{figure}

From here we now return to the physics of the Gnevyshev gap. The answer is 
quite simple. The Gnevyshev intervals in the butterfly diagrams are related to 
the 11-year modulations of the ADM density, trapped on the Sun (via vertical 
density waves from the disk to the solar neighborhood) and, as a result, to 
anticorrelated indicators of variations in the sunspot cycles or photons of 
axion origin. When the ADM density in the solar interior is low (or 
the sunspot count is high), then the Gnevyshev intervals will be short
in time. Conversely, when the ADM density in the solar interior is high (or the
sunspot count is low), then the Gnevyshev intervals will be longer in time. As 
no surprisingly, it is the presence of ADM density modulations that shows that 
the appearance of ``our indicated'' Gnevychev gap is the result of precisely 
11-year variations of butterfly cycles with sunspots.

The most interesting thing is that it is the 11-year variations of sunspot 
cycles or photons of axion origin (see Figs.~5a and~6b 
in~\cite{RusovDarkUniverse2021}) that are anticorrelated not only with the 11-year
modulations of ADM density on the Sun, but also with the 11-year modulations of
ADM density in the solar neighborhood, and, in particular, on Earth (see e.g. 
Fig.~\ref{fig-time-variations}).

First, we recall that the physics of experimental data related to the 
anticorrelation between 11-year variations in the luminosity of the Sun and the
Earth (see Fig.~\ref{fig-time-variations}) currently cannot explain one of the
most mystical phenomena of solar-terrestrial physics, which is contained in 
well-known experiments, for example, between the temporal variations of the 
magnetic flux of the solar tachocline (black curve; see~\cite{Dikpati2008ApJ675} and Fig.~16 in~\cite{Rusov2013}), and secular 
variations in the Earth's magnetic field (Y-component from~\cite{Geomagnetism2025}), or variations in the Earth's rotation
speed (see~\cite{Sidorenkov2016,Sidorenkov2009}
),
as well as the variations in the mean global sea 
level (Pacific Decadal Oscillation (PDO)~\cite{PDO2025} + Atlantic Multidecadal Oscillation 
(AMO)~\cite{AMO2023}) 
with a trend in mm/years near 60-year oscillation (see Fig.~1 in~\cite{Chambers2012}),
between which there is a time lag of about 10-11~years, respectively (see Fig.~\ref{fig-time-variations}; and also Fig.~6 in~\cite{RusovDarkUniverse2021}.

%

\section{Summary and Outlook}
\label{sec-summary}

Most importantly, the existing void in the tube causes high velocities and 
short lifetimes of sunspots from the tachocline to the photosphere, which is
consistent with experimental data. This means that the modernity of a 
well-developed helioseismological inversion, surprisingly, persists to this 
day.

The perspective is to understand the physics of the anticorrelation between 
11-year variations of total solar irradiance and 11-year variations of the 
magnetic field and other Earth parameters (see Fig.~\ref{fig-time-variations}).

\bibliography{Rusov-AxionSunLuminosity}

\end{document}